\documentclass[prd,aps,superscriptaddress,nofootinbib,showpacs,twocolumn,
preprintnumbers]{revtex4}

\usepackage[intlimits]{amsmath}
\usepackage{amsfonts}
\usepackage{amssymb,amscd}
\usepackage{bbold}
\usepackage{color}
\usepackage{graphics}
\usepackage[dvips]{epsfig}

\newcommand{\beq}{\begin{equation}}
\newcommand{\eeq}{\end{equation}}
\newcommand{\beqa}{\begin{eqnarray}}
\newcommand{\eeqa}{\end{eqnarray}}
\newcommand{\be}{\begin{equation}}
\newcommand{\ee}{\end{equation}}

\def\qq{\text{\boldmath$q$\unboldmath}}

\def\pp{\text{\boldmath$p$\unboldmath}}

\def\kk{\text{\boldmath$k$\unboldmath}}
\def\bmu{\text{\boldmath$\mu$\unboldmath}}

\def\gbold{\text{\boldmath$\gamma$\unboldmath}}

\def\vpp{\vert \pp \vert}
\def\vqq{\vert \qq \vert}

\newcommand*{\mn}{{\mu\nu}}

\begin{document}
\preprint{INT-PUB-10-043}

\title{Quark spectral properties above $T_c$ from Dyson-Schwinger equations}
\author{Jens~A.~Mueller}
\affiliation{Institut f\"ur Kernphysik, 
  Technische Universit\"at Darmstadt,
  Schlossgartenstra{\ss}e 9,\\ 
  D-64289 Darmstadt, Germany}
\author{Christian~S.~Fischer}
\affiliation{Institut f\"ur Kernphysik, 
  Technische Universit\"at Darmstadt,
  Schlossgartenstra{\ss}e 9,\\ 
  D-64289 Darmstadt, Germany}
\affiliation{Institut f\"ur Theoretische Physik, 
 Universit\"at Giessen, 35392 Giessen, Germany}
\affiliation{GSI Helmholtzzentrum f\"ur Schwerionenforschung GmbH, 
  Planckstr. 1  D-64291 Darmstadt, Germany.}
\author{Dominik~Nickel}
\affiliation{Institute for Nuclear Theory, University of Washington, Seattle, WA 98195}

\date{\today}
\begin{abstract}
We report on an analysis of the quark spectral representation at finite temperatures based on
the quark propagator determined from its Dyson-Schwinger equation in Landau gauge.
In Euclidean space we achieve nice agreement with recent results
from quenched lattice QCD. We find different analytical properties of the 
quark propagator below and above the deconfinement transition.
Using a variety of ans\"atze for the spectral function we then analyze
the possible quasiparticle spectrum, in particular its quark mass and
momentum dependence in the high temperature phase.
This analysis is completed by an application of the Maximum Entropy Method, in
principle allowing for any positive semi-definite spectral function.
Our results motivate a more direct determination of the spectral function in
the framework of Dyson-Schwinger equations.
\end{abstract}

\maketitle

\section{introduction}

The wealth of data produced by the Relativistic Heavy-Ion Collider (RHIC)
 has lead to new insights about the nature of the quark-gluon plasma (QGP),
 in particular in the strongly coupled regime just above the chiral phase
 transition (see Refs.~\cite{Muller:2006ee,Shuryak:2008eq,BraunMunzinger:2009zz,Heinz:2009xj}
 for recent reviews). Despite the strong coupling, phenomena like the constituent
 quark number scaling of elliptic flow~\cite{Molnar:2003ff} still suggest quarks
 as quasiparticle excitations. This should also be connected with the success
of quasiparticle models in describing thermodynamic properties in this regime 
of the QCD phase diagram~\cite{Schulze:2007ac,Cassing:2007yg} and also serves
 as a foundation for transport approaches like the one discussed in Ref.~\cite{Cassing:2008sv}.
In another context, dilepton production in a heavy ion collision has been
related to the spectral properties of the thermalized quasiparticles and specifically
to the dispersion relation of quarks~\cite{Braaten:1990wp,Peshier:1999dt,Arnold:2002ja}.
We therefore conclude that a detailed understanding of the quasiparticle spectrum, 
in particular close to the chiral phase transition, is desirable.

Our quantity of interest is the quark spectral function, which
 encodes all information about the two-point correlation function, in particular
 also its poles and consequently the quark's dispersion relation and decay width.
In the realm of weak coupling reliable results are obtained using the hard-thermal
loop (HTL) expansion \cite{Braaten:1989mz,Baym:1992eu,Blaizot:1993bb}:
One finds the by now well-known pattern of two quasiparticle excitations and a
 continuum contribution stemming from a branch cut in the quark propagator due to
Landau damping, i.e. the absorption of a space-like quark by a hard gluon or hard antiquark.
The two quasiparticles correspond to an ordinary quark with a positive ratio of
chirality to helicity and to a collective mode dubbed 'plasmino' with a negative
chirality to helicity ratio. Both have thermal masses of order $gT$ and decay
widths of order $g^2T$, where $g$ is the coupling constant and $T$ the temperature.

Beyond systematic weak-coupling expansions, there is no straightforward approach
 for the determination of the spectral function or its characteristics.
Some insight has been gained using model calculations, finding similar structures
 as in the weak coupling limit besides a possible additional third quasiparticle 
excitation which is only present at small momenta
\cite{Schaefer:1998wd,Kitazawa:2005mp,Kitazawa:2006zi}.
The width of the quasiparticle peaks, which is of order $g^2T$ in HTL approximation,
 might however become of the order of the thermal mass for
couplings at the scale of the chiral phase transition~\cite{Harada:2007gg}.
This questions the existence of well-defined quasiparticles at least at small momenta.

Ab initio calculations of correlators within lattice QCD 
are performed in Euclidean space and an analytic continuation
that is necessary to determine the spectral function is strictly
speaking not possible.
There are, however, approaches that aim to extract attributes of
spectral functions from the numerical data.
One of these is the Maximum Entropy Method
(MEM)~\cite{Jarrell:1996,nr:1997}, which has been applied in similar
contexts, e.g. for extracting the spectral functions of mesons~\cite{Asakawa:2000tr,Karsch:2001uw}.
For cold and dense matter MEM has also been applied within the framework
of Dyson-Schwinger equations (DSEs) in Ref.~\cite{Nickel:2006mm}.
In some sense MEM corresponds to fitting the ``most likely'' 
spectral function to the data but without restricting the form 
of the spectral function. 
In contrast to this another method has been established in Refs.~\cite{Karsch:2007wc,Karsch:2009tp}.
There the authors assumed a certain shape for the quark spectral function
including a few fitting parameters which were determined from the numerical data.
In such an approach one relies on physical
guidance as e.g. given by the HTL approximation for the construction
of a suitable ansatz for the spectral function.

A complementary approach to the correlators of QCD are functional methods 
such as Dyson-Schwinger equations and the functional renormalisation group. 
In recent years, these methods have been successfully applied to problems such 
as the characterization of the chiral and deconfinement transition 
\cite{Braun:2007bx,Braun:2008pi,Marhauser:2008fz,Fischer:2009wc,Fischer:2009gk,Braun:2009gm,Fischer:2010fx},
or the properties of gluons in the high temperature phase 
\cite{Maas:2004se,Maas:2005hs,Cucchieri:2007ta}. 
In this work we do the first steps towards the determination of spectral functions 
in the framework of Dyson-Schwinger equations beyond simple rainbow approximations.
Particularly due to our input for the gluon propagator the DSE calculations are also 
limited to Euclidean
space and we use the two methods described above to explore the possible shape of the
spectral function.
The truncation scheme is detailed in section \ref{sec:DSE} and has been 
shown to reproduce the chiral and deconfinement transition 
temperatures of quenched lattice QCD in Ref.~\cite{Fischer:2010fx}.
In section \ref{sec:prop} we briefly present our results for the
quark dressing functions in Euclidean space below and above the
critical temperature $T_c$.
We then introduce the spectral representation for the quark propagator in section~\ref{sec:spectral_rep}.
Using fit ans\"atze for the spectral function, we focus on zero momentum and finite quark masses in
section~\ref{sec:prop_zero_mom} before moving to finite momentum in the chirally restored phase in section~\ref{sec:mom_dep}.
The obtained results are confronted with recent quenched lattice QCD results from Ref.~\cite{Karsch:2009tp}.
Finally we evaluate our truncation scheme using MEM in section~\ref{sec:MEM}, before we conclude in section \ref{sec:sum} and
outline possible future directions.

\section{quark Dyson-Schwinger equation and employed
  truncations \label{sec:DSE}}

\subsection{Quark Dyson-Schwinger equation}
The renormalized quark Dyson-Schwinger equation in the Matsubara
formalism is given by
\begin{align}
 S^{-1}(i\omega_p,\pp)
&=
Z_2\,S_0^{-1}(i\omega_p,\pp)-\Sigma(i\omega_p,\pp)
\,,
\label{quark:dse}
\end{align}
with the inverse full quark propagator $S^{-1}(i\omega_p,\pp)$, the quark
wave function renormalization constant $Z_2$,
the inverse bare quark propagator $S^{-1}_{0}(i\omega_p,\pp)$
and the to-be-specified self-energy $\Sigma(i\omega_p,\pp)$.
With the only difference of explicitly using imaginary arguments for the
energy in all functions, we follow the conventions from
Ref.~\cite{Fischer:2010fx}.
For the fermionic Matsubara frequencies we have $\omega_p=
(2n_p+1)\pi T$ with temperature $T$.
For the Dirac structure of the inverse propagators we have
\begin{align} 
S^{-1}_0(i\omega_p,\pp)
&=
i\gamma_4\,\omega_p
+
i\text{\boldmath{$\gamma$}\unboldmath}\cdot\pp
+
Z_m\,m(\mu)
\,,\\
S^{-1}(i\omega_p,\pp)
&=
i\gamma_4\omega_p C(i \omega_p,\vert\pp\vert)
+
i\text{\boldmath{$\gamma$}\unboldmath}\cdot\pp\,A(i \omega_p,\vert\pp\vert)
\nonumber\\
&\phantom{=}
+
B(i \omega_p,\vert\pp\vert)
\,,
\label{Dirac:decomp}
\end{align}
i.e. the inverse full propagator is para\-meterized by the dressing
functions $A(i \omega_p,\vert\pp\vert)$, $B(i\omega_p,\vert\pp\vert)$
and $C(i \omega_p,\vert\pp\vert)$.
The quark mass renormalization constant $Z_m$ and the renormalized
mass $m(\mu^2)$ at renormalization scale $\mu$, together with $Z_2$,
are then determined within a MOM-renormalization scheme.

In Landau gauge, the self-energy takes the form
\begin{align}\nonumber
\Sigma(i\omega_p,\pp)
&=
\frac{16\pi}{3}\dfrac{Z_2}{\tilde{Z}_3} \alpha(\mu)
T\sum_{n_q}\int\frac{d^3q}{(2\pi)^3}\;\Big[\gamma_{\mu}S(i\omega_{q},\qq)
\\&
\times\Gamma_{\nu}(i\omega_q,\qq,i\omega_p,\pp)D_{\mu\nu}(i\omega_p-i\omega_q,\pp-\qq)\Big]
\,,
\label{quark:self-e}
\end{align} 
with gluon propagator $D_{\mu\nu}$, quark-gluon vertex $\Gamma_{\nu}$,
gauge coupling $\alpha(\mu)$ and ghost renormalization constant
$\tilde{Z}_3$.
Assuming we know $D_{\mu\nu}$, $\Gamma_{\nu}$ and the renormalization
procedure, the DSE then translates into coupled non-linear integral
equations for the desired dressing functions.

\begin{figure*}[t]
\includegraphics[width=1.\columnwidth]{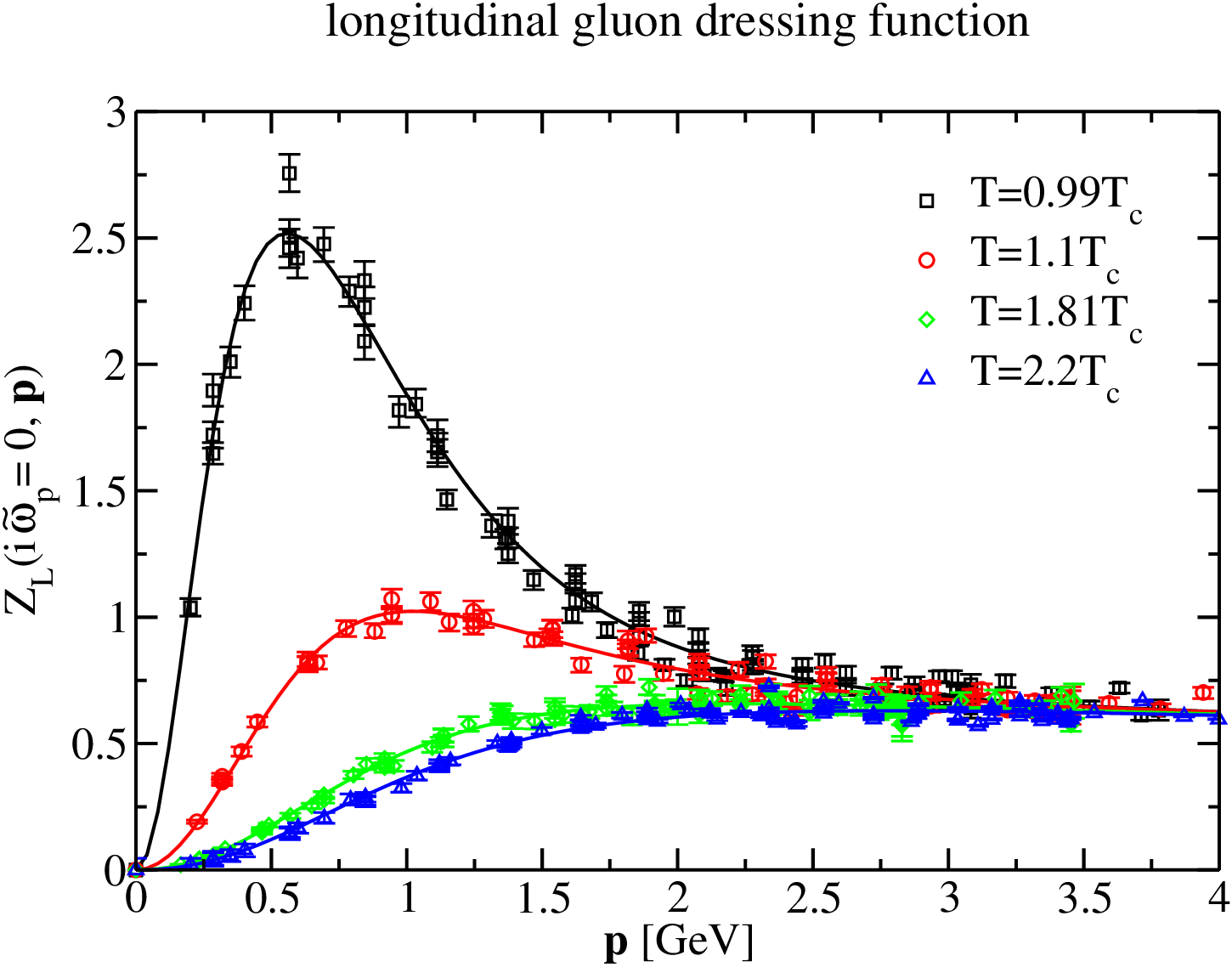}\hfill
\includegraphics[width=1.\columnwidth]{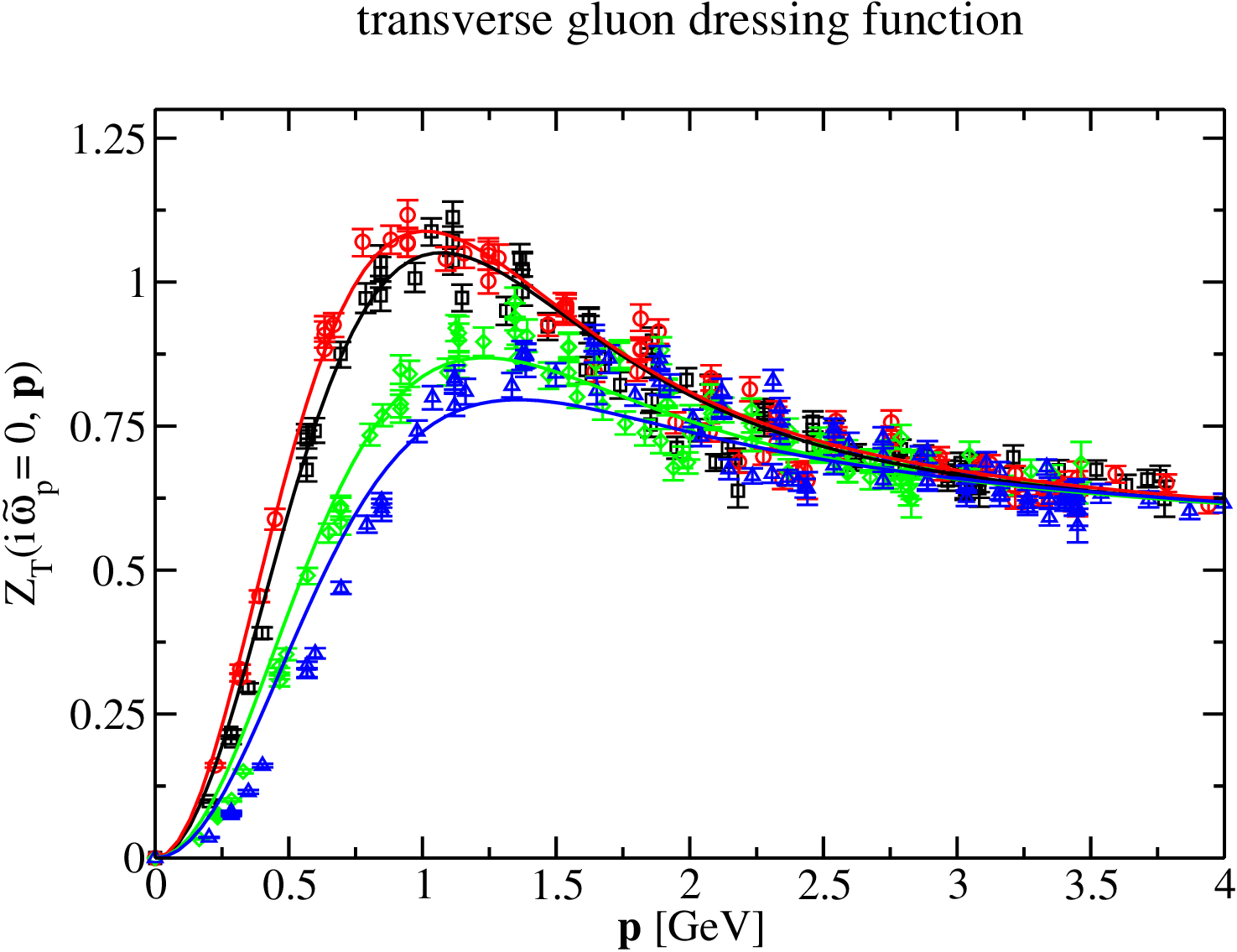}
\caption{Longitudinal (left) and transverse (right) temperature dependent gluon
dressing function for the lowest Matusbara frequency from lattice calculations 
compared with our fits. Shown are results for temperatures below the 
critical temperature ($0.99\,T_c$) and above the critical temperature 
($1.1\,T_c,\;1.81\,T_c,\;2.2\,T_c$).
}
\label{res:fitscompared}
\end{figure*}

\subsection{Truncation setup}
The gluon propagator being a major ingredient in the quark DSE,
can in principle be obtained in quenched approximation from
corresponding DSE's \cite{Cucchieri:2007ta,Lichtenegger:2008mh,Maas:2005hs,Maas:2004se}.
However, this turns out to be a formidable task and the results are
still on a qualitative level only. We therefore use results from 
lattice gauge theory as input. These have
been made available recently on a fine temperature grid over a
temperature range from $0$ to $2.2\;T_c$ \cite{Fischer:2010fx}.

In Landau gauge the gluon propagator is given through transverse and
longitudinal dressing functions $Z_T(i\tilde{\omega}_k,\vert\kk\vert)$
and $Z_L(i\tilde{\omega}_k,\vert\kk\vert)$, respectively, via
\begin{align}
D_{\mu\nu}(i\tilde{\omega}_k,\kk)
&=
\frac{Z_T(i\tilde{\omega}_k,\vert\kk\vert)}{k^2}P^T_{\mu\nu}(k)
+
\frac{Z_L(i\tilde{\omega}_k,\vert\kk\vert)}{k^2}P^L_{\mu\nu}(k)
\,.
\label{glue:prop}
\end{align}
Hereby we introduced the projectors
\begin{align}
P_\mn^T(k)
&=
(1-\delta_{\mu4})(1-\delta_{\nu4})\left(\delta_\mn-\frac{k_\mu
    k_\nu}{\kk^2}\right)
\,,
\nonumber\\
P_\mn^L(k) &=\delta_\mn-\frac{k_\mu k_\nu}{k^2}-P_\mn^T(k)
\,,
\end{align}
the bosonic Matsubara frequency $\tilde{\omega}_k=2\pi n_k T$ and the
short-hand notation $k_\mu=(\tilde{\omega}_k,\kk)_\mu$, for which
$k^2=\tilde{\omega}_k^2 + \kk^2$.
As in the case of quark propagator and quark-gluon vertex we suppress
color indices due to the remaining global color symmetry.

The gluon dressing functions, $Z_T$ and $Z_L$, have been calculated in quenched lattice
gauge theory for gauge group SU(2) and SU(3) \cite{Fischer:2010fx}.
In the quark Dyson-Schwinger equation we also need to evaluate the
gluon propagator for momenta not identically to the ones of the
lattice calculation. Therefore we use temperature dependent fits to
these data.
In Fig.~\ref{res:fitscompared} we show results for gauge group
SU(3) which is the relevant gauge group in this work. 
The figure presents the lattice results denoted by data points together with 
corresponding fit functions represented by straight lines for a variety
of temperatures.
The explicit expressions for the fit functions and a more detailed
discussion concerning the gluon propagator can be found in
Ref.~\cite{Fischer:2010fx} and shall not be repeated here for
brevity.

The details of the second ingredient for a closed quark DSE, namely the
quark-gluon vertex, are yet to be explored. First exploratory results
on the mass and momentum dependence of the vertex at zero temperature
have been reported from lattice calculations and Dyson-Schwinger
equations, see \cite{Kizilersu:2006et,Alkofer:2008tt} and
Refs. therein. However, not much is known about its temperature
dependence. In such a situation a viable strategy is to use
phenomenological model ans{\"a}tze for the vertex which are then
justified by comparing results with other approaches. This strategy
has been successful in previous works \cite{Fischer:2010fx} and will therefore also be adopted here.

In the following we employ the
temperature dependent ansatz for the vertex used in Refs.
\cite{Fischer:2009wc,Fischer:2009gk,Fischer:2010fx}. It is given by
\begin{widetext}
\beqa \label{vertexfit}\nonumber
\Gamma_\nu(i\omega_q,\qq,i\omega_p,\pp) = \widetilde{Z}_{3}\left(\delta_{4 \nu} \gamma_4 
\frac{C(i\omega_q,\vqq)+C(i\omega_p,\vpp)}{2}
+  (1-\delta_{4 \nu}) \gamma_\nu 
\frac{A(i\omega_q,\vqq)+A(i\omega_p,\vpp)}{2}
\right)\\ [.2cm]
\times\left( 							
\frac{d_1}{d_2+k^2} 			
 + \frac{k^2}{\Lambda^2+k^2}
\left(\frac{\beta_0
    \alpha(\mu)\ln[k^2/\Lambda^2+1]}{4\pi}\right)^{2\delta}\right)
\,,
\label{eq:vertex}
\eeqa 
\end{widetext}
with the gluon momentum $k_\mu=(\omega_p-\omega_q,\pp-\qq)_\mu$ and $\beta_0=11$ for quenched QCD.
The anomalous dimension of the vertex $\delta$ accounts for 
the correct perturbative running coupling. It is given by 
$\delta=-9/44$ and we renormalize at $\alpha(\mu)=0.3$.
The Dirac structure with the quark vector dressing functions
$A,\;C$ represents the first component of the Ball-Chiu vertex
at finite temperature. It is motivated by the Slavnov-Taylor
identity for the vertex and constitutes a first approach
towards a more realistic non-trivial temperature dependence of
the vertex. In the additional phenomenological fit function we
have parameters $d_1$ and $d_2$ and the temperature
independent scale $\Lambda=1.4$ $\rm{GeV}$. 
Since we use lattice results as reference for the gluon propagator, the
quark-gluon vertex is the only uncertainty source in our calculation.
We adjust the parameter $d_1$ temperature dependent while
keeping $d_2=0.5$ $\rm{GeV}^2$ independent of temperature. As for $d_1$ it turns 
out that its temperature dependence is in one-to-one correlation 
to the thermal masses $m_T$ of the quarks as extracted in
section~\ref{sec:prop_zero_mom}.
Since the same quantity has been determined from quenched QCD results
for the Landau gauge quark propagator, see Tab.~II of
Ref.~\cite{Karsch:2009tp}, we adjust $d_1$ such that it roughly coincides
with these results.
This procedure works only above the critical temperature where 
\begin{table}[b]
\caption{Temperature dependent parameter for the quark-gluon vertex
  and extracted thermal masses $m_T$.\label{tab1}}
\begin{tabular}{c||c|c|c|c}
$T/T_c$    &  $<1$ & 1.25& 1.5  & 2.2 \\\hline\hline
$d_1$ $[{\rm GeV}^2]$    & 4.6 & 0.5& 0.4  & 0.25\\   
$m_T/T$    &  & 0.865 & 0.862 & 0.822\\\hline
\end{tabular}
\end{table}
thermal masses can be extracted. Below the critical temperature no
such information is available and we therefore choose the constant
value $d_1=4.6$ $\rm{GeV}^2$. This choice ensures spontaneous chiral symmetry
breaking and agrees with the one used in
Ref.~\cite{Fischer:2010fx}.
We certainly checked the $d_1$ dependence of our results for the quark
spectral function below the critical temperatures; this is discussed
in section \ref{sec:prop_zero_mom}.
Our values of $d_1$ and the obtained  thermal masses are summarized in
Tab.~\ref{tab1}.
Obviously, the strength of the vertex ansatz at low momenta is
significantly reduced above the critical temperature.
This is what one may expect: In the Dyson-Schwinger equation
for the quark-gluon vertex, discussed in detail in
Ref.~\cite{Alkofer:2008tt}, we find skeleton diagrams containing
dressed gluon propagators.
As is evident from  Fig.~\ref{res:fitscompared} there is a rapid
decrease in the strength of the longitudinal gluon at or around the
critical temperature $T_c$.
This rapid decrease will backfeed into the vertex DSE which probably explains the
reduction of the strength of the quark-gluon vertex.
This  behavior fits nicely to our findings for the parameter
$d_1$ in our ansatz for the vertex.
We therefore believe that our vertex ansatz accurately reflects
qualitative and maybe even quantitative features of the fully dressed
quark-gluon vertex.

\section{Matsubara propagator below and above $T_c$} \label{sec:prop}

\begin{figure}
\includegraphics[width=.9 \columnwidth]{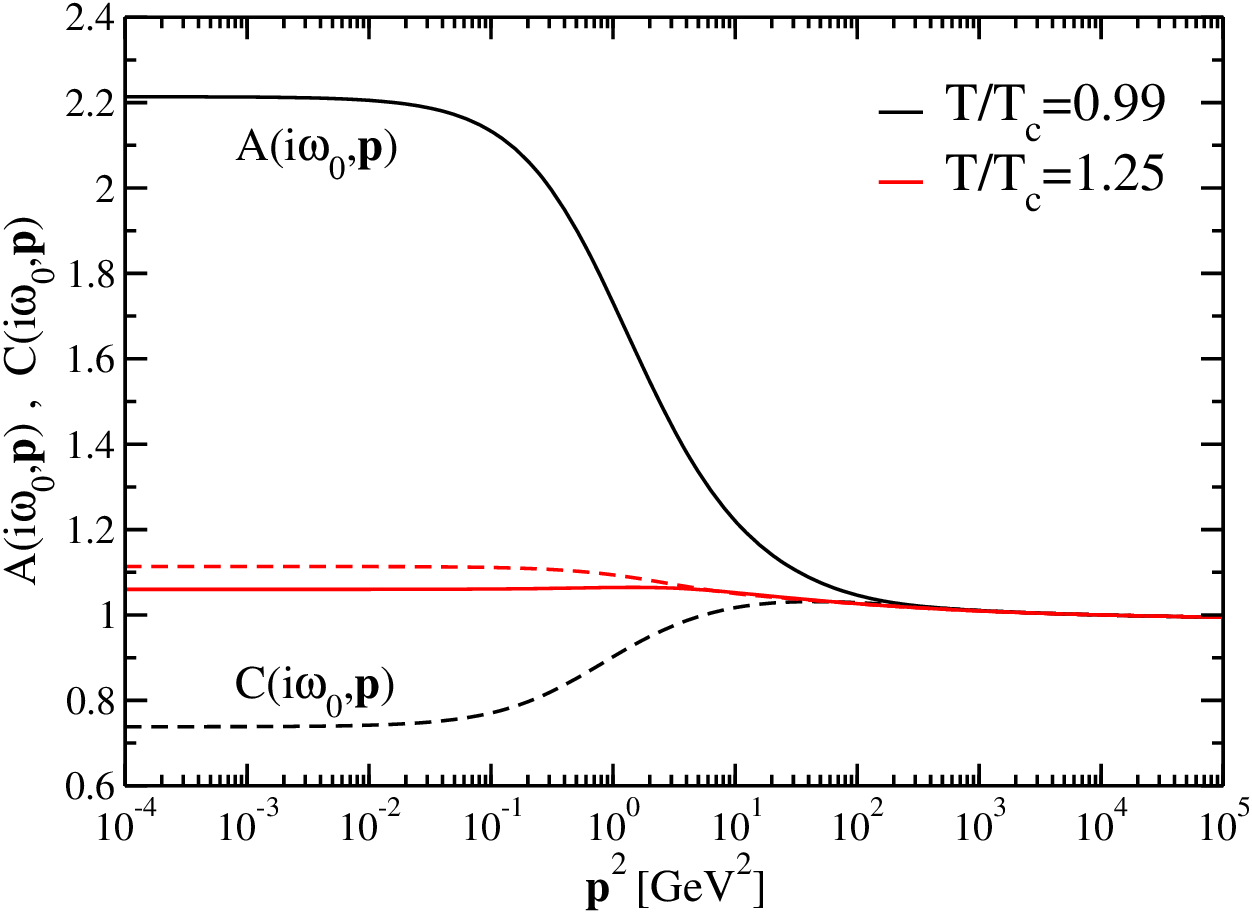}\\\vspace*{4mm}
\includegraphics[width=.9 \columnwidth]{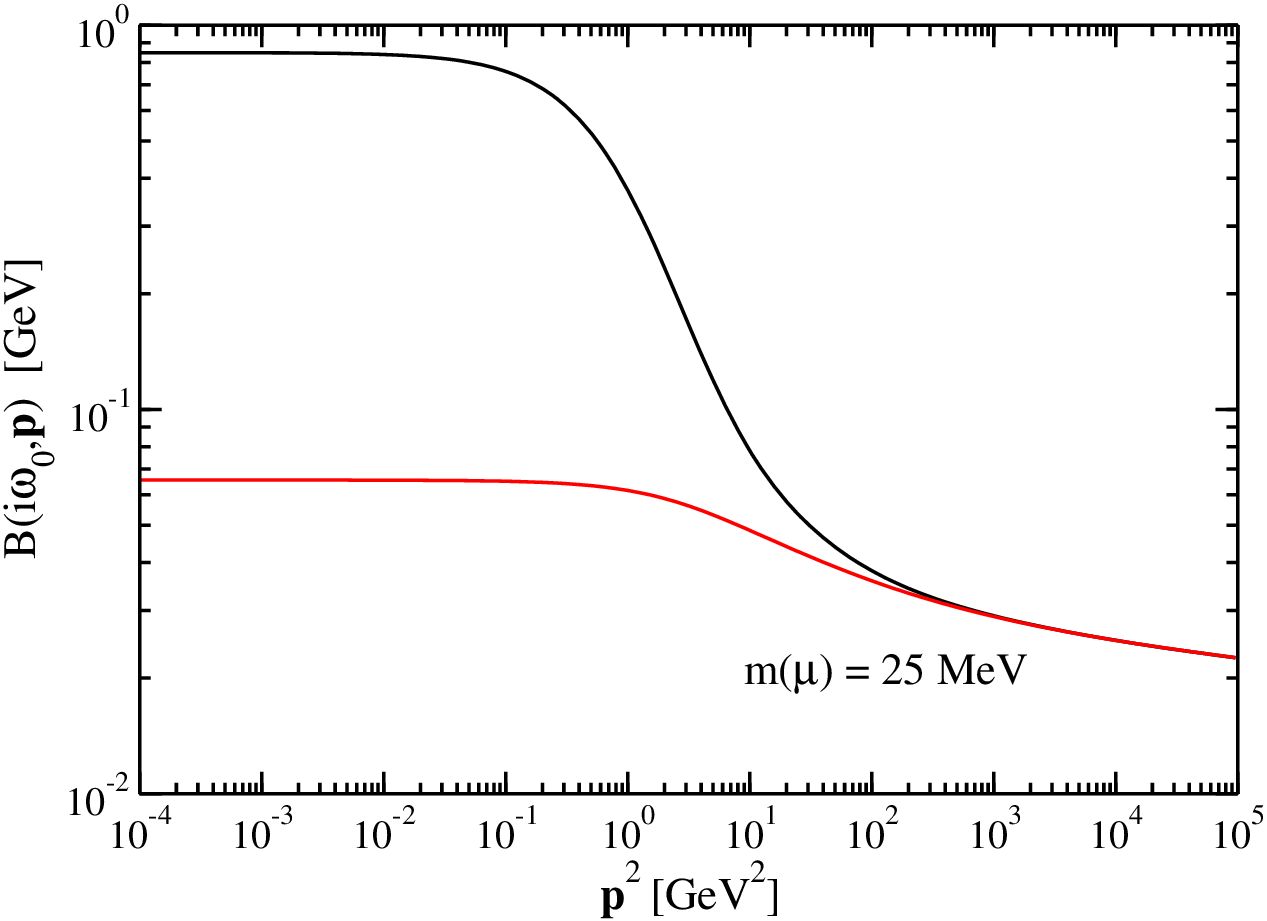}
\caption{Momentum dependence of the dressing functions $A,\,C,\,B$ below and 
above the critical temperature. Results are shown for the lowest Matsubara frequency
and finite bare quark mass $m(\mu)=25\,\text{MeV}$.\label{fig:prop}
}
\end{figure}

Before we focus on the spectral functions we briefly discuss our results for 
the Matsubara propagator in momentum space shown in Fig.~\ref{fig:prop}.
The results are obtained employing a sharp cutoff $\Lambda$
for the computation of the self-energy.
Thus the integration and summation extends to momenta and 
frequencies with $\omega_q^2 + \qq^2 \leq \Lambda^2$.
For $\vert n_q\vert \leq 39$ the summation over Matsubara frequencies is performed
explicitly. The remaining sum is approximated by an integral and we checked
that the results are insensitive to a change of the number of explicitly summed
Matsubara modes. 
We then use a MOM-renormalization scheme with renormalization conditions $C(i\omega_{0},\bmu)=1$
and $B(i\omega_0,\bmu)=m(\mu)$ and $\omega^2_{0}+\bmu^2=10545$ $\rm{GeV}^2$;
consequently all dressing functions are independent of the cutoff $\Lambda$.
More details concerning the numerical procedure can be found in Ref.~\cite{Fischer:2009gk}.

In Fig.~\ref{fig:prop} we show results for $m(\mu)=25$ $\rm{MeV}$.
Above $T_c$ the interaction strength as parameterized by the
quark-gluon vertex is strongly reduced and the dressing functions $A(i\omega_0, \pp)$ and $C(i\omega_0,\pp)$,
which are finite in Landau gauge,  do not deviate much from unity.
In the infrared all dressing functions are constant and
we observe that $A(i\omega_0, \pp)<C(i\omega_0,\pp)$.
This is also found for the dressing functions in HTL approximation~\cite{Bellac:1996}, where this inequality holds for all momenta.
In the simplified case of constant dressing functions this is
perfectly consistent with our expectations:
As can be seen from Eq.~(\ref{Dirac:decomp}), we would interpret
$\sqrt{A/C}=v\leq1$ as the velocity of the quasiparticles, $\sqrt{v^2
  \pp^2+B^2/C^2}$ as their dispersion relation and $1/C\leq1$ as the
wave-function renormalization of the
$\gamma_4$-component\footnote{Note that only the $\gamma_4$-component of the quark propagator obeys a sum rule \cite{Nickel:2006mm}.}.
In contrast to the HTL approximation there is, however,  a momentum range
 ($8\text{ GeV}^2$ to $200 \text{ GeV}^2$ in Fig.~\ref{fig:prop})
 where $A(i\omega_0,\pp)>C(i\omega_0,\pp)$
for all here considered temperatures above $T_c$ and all masses.
To what extent this behavior is generic or truncation-affected needs
to be investigated in future studies.

Below $T_c$, on the contrary, the results become less intuitive. Due
to the stronger interaction and lower temperature, the chiral symmetry
breaking dressing function $B$ is significantly larger in the infrared
with its size related to the scale of dynamical symmetry breaking.
For the dressing functions $A(i\omega_0, \pp)$ and $C(i\omega_0,\pp)$ we
find, however, $A(i\omega_0, \pp)>C(i\omega_0,\pp)$. This does not allow
for a simple quasiparticle interpretation as outlined above.
Also the contribution from the self-energy is of the same order as the
free propagator, in particular the dressing function $A(i\omega_0,
\pp)$ is strongly enhanced in the infrared.
This also hints towards more complicated dynamics than a gas of
quasiparticles, which is of course expected in the confined phase of QCD.

From these results we therefore conclude that a simple interpretation
in terms of the excitation spectrum seems not appropriate below $T_c$. Even
for the analysis above $T_c$, as we discuss in the following, the
quasiparticles can receive a thermal mass in the chirally restored
phase. This is already known from the HTL expansion, where the
quasiparticles in the high temperature and small coupling regime
have a mass of order $gT$.

\section{Quark spectral functions and representation}\label{sec:spectral_rep}

Causality and resulting analyticity imply that a correlator can be
represented through a spectral function.
For the quark propagator this representation is given by
\begin{align}
S(i\omega_n,\pp)
&=
\int_{-\infty}^\infty
\!\frac{d\omega'}{2\pi}\,
\frac{\rho(\omega',\pp)}{i\omega_n-\omega'}
\,,
\end{align}
where the Dirac structure of the spectral function is parameterized as
\begin{align}
\rho(\omega,\pp)
&=
2\pi\big(
\rho_{4}(\omega,\vert\pp\vert)\gamma_4
+
\rho_{\rm v}(\omega,\vert\pp\vert)\left( i\gbold\cdot\pp\right)/\vert\pp\vert
\nonumber\\&\phantom{=2\pi\big(}
-
\rho_{\rm s}(\omega,\vert\pp\vert)
\big)
\,.
\end{align}
Our conventions are chosen such that the scalar dressing functions themselves
agree with those introduced in Ref.~\cite{Karsch:2009tp} using
Minkowski space conventions\footnote{
$\gamma^\mu_{\rm M}$ in common Minkowski space conventions are related
to those used in this work via $\gamma_{\rm M}^0=-\gamma_4$ and
$\gamma_j=-i\gamma^j_{\rm M}$.
}.
Assuming a positive definite Fock space\footnote{For completeness we
  note that this is not guaranteed for a gauge-fixed Yang-Mills
  theory.}, the dressing functions furthermore obey
\begin{align}
\rho_{4}(\omega,\vert\pp\vert)
&\geq
\sqrt{
\rho_{\rm v}(\omega,\vert\pp\vert)^2+\rho_{\rm s}(\omega,\vert\pp\vert)^2
}
\geq
0
\label{eq:RhoInequality}
\end{align}
and the sum rules
\begin{align}
1&=Z_2\int_{-\infty}^\infty\!d\omega\, \rho_4(\omega,\vert\pp\vert)
\,,\label{eq:sum_rule1}\\
0&=\phantom{Z_2}\int_{-\infty}^\infty\!d\omega\, \rho_{\rm v}(\omega,\vert\pp\vert)
\,,\label{eq:sum_rule2}\\
0&=\phantom{Z_2}\int_{-\infty}^\infty\!d\omega\, \rho_{\rm s}(\omega,\vert\pp\vert)
\,,
\label{eq:RhoSumRules}
\end{align}
where here $Z_2$ is the wave function renormalization constant and not
the plasmino residue which will be introduced later.

In the following we limit ourself to vanishing momenta $\vert\pp\vert$
and the chirally restored phase, respectively.
For this purpose it is instructive to introduce the projectors
\begin{align}
P_\pm(\pp)
&=
\frac{1}{2}\left(
1\mp i\gamma_4 \gbold\cdot\pp/\vert\pp\vert
\right)
\,,\nonumber\\
L_\pm
&=
\frac{1}{2}\left(
1\mp\gamma_4
\right)
\,,
\end{align}
where the signs are again chosen in order to agree with common
Minkowski space conventions. $P_\pm(\pp)$ can be
interpreted as energy projectors for massless modes.

For vanishing momentum it is then convenient to write
\begin{align}
\rho(\omega,{\mathbf{0}})
&=
\rho^M_+(\omega)L_+\gamma_4
+
\rho^M_-(\omega)L_-\gamma_4
\,,\\
S(i\omega_n,{\mathbf{0}})
&=
S^M_+(i\omega_n)L_+\gamma_4
+
S^M_-(i\omega_n)L_-\gamma_4
\,,
\end{align}
i.e.
$\rho^M_\pm(\omega)=2\pi\big(\rho_4(\omega,{\mathbf{0}})\pm\rho_{\rm s}(\omega,{\mathbf{0}})\big)$
and
$\rho_{\rm v}(\omega,{\mathbf{0}})=0$.
From Eqs.~(\ref{eq:RhoInequality}) and (\ref{eq:RhoSumRules}) we infer
that the functions $\rho^M_\pm(\omega)$ are positive semi-definite and
normalized to $2\pi Z_2$. Furthermore the spectral representation yields
the scalar relation
\begin{align}
S^M_{\pm}(i\omega_n)
&=
\int_{-\infty}^{\infty}\!\frac{d
  \omega'}{2\pi}\,\frac{\rho^M_{\pm}(\omega')}{i\omega_n-\omega'}
\,.
\label{prop:spectral}
\end{align}

On the other hand for a chirally symmetric phase with
$B(i\omega,\vert\pp\vert)=0$ and $\rho_{\rm s}(\omega,\vert\pp\vert)=0$, we
introduce
\begin{align}
\label{eq:rhoP}
\rho(\omega,\pp)
&=
\sum_{e=\pm}
\rho^P_{e}(\omega,\vert\pp\vert)
\,P_e(\pp)\,
\gamma_4
\,,\\
\label{eq:SP}
S(i\omega_n,\pp)
&=
\sum_{e=\pm}
S^P_{e}(\omega,\vert\pp\vert)
\,P_e(\pp)\,
\gamma_4
\,,
\end{align}
i.e. $\rho^P_\pm(\omega,\vert\pp\vert)=2\pi\big(\rho_{4}(\omega,\vert\pp\vert)\pm\rho_{\rm
  v}(\omega,\vert\pp\vert)\big)$.
As before we see that $\rho^P_\pm(\omega,\vert\pp\vert)$ is positive
semi-definite and normalized to $2\pi Z_2$. The spectral representation for
the dressing functions again take the form
\begin{align}
S^P_\pm(i\omega_n,\vert\pp\vert)
&=
\int_{-\infty}^{\infty}\!\frac{d\omega'}{2\pi}\,
\frac{\rho^P_\pm(\omega',\vert\pp\vert)}{i\omega_n-\omega'}
\label{prop:spectralP}
\,.
\end{align}

In the following we aim to invert the linear relation between
Matsubara propagators $S^M_\pm(i\omega_n)$, $S^P_\pm(i\omega_n,\pp)$
determined by solving the truncated DSE and the respective spectral
functions $\rho^M_{\pm}(\omega)$ and
$\rho^P_\pm(\omega,\vert\pp\vert)$.
Strictly speaking, for a finite set of Matsubara frequencies, this
problem is however ill-posed.

\section{Quark spectral functions at zero
  momentum} \label{sec:prop_zero_mom}

In this section we focus our attention to the quark propagator
at zero momentum. We analyze the correlator at temperatures
below and above the critical temperature and for various bare
quark masses.
Since the general problem of extracting the spectral function from the
spectral representation is ill-posed, we will limit ourselves to
parameterized ans\"atze and determine the best fitting function in the
given subspace.
Above the critical temperature we study the quark mass dependence of
the so obtained insight on quasiparticle excitations. This strategy
was explored in Refs.\cite{Karsch:2007wc, Karsch:2009tp} in the
framework of lattice QCD. 

We will mainly consider the following two-pole ansatz for the spectral
function
\begin{align}
\rho^{M}_\pm(\omega)&=2\pi\big[Z_1\delta(\omega\mp E_1)+Z_2\delta(\omega\pm E_2)\big]
\,,
\label{two_pole:delta}
\end{align}
with fitting parameters $Z_1,\;Z_2$ and $E_1,\;E_2$.
This will be related to our quark propagator via
Eq.~(\ref{prop:spectral}), for which we identify
\begin{align}
S^{{\rm DSE},\,M}_{\pm}(i\omega_n)
&=
-
\frac{
i\omega_n\,C(i\omega_n,0) \pm B(i\omega_n,0)
}{
\omega_n^2\,C^2(i\omega_n,0)+B^2(i\omega_n,0)
}
\,.
\label{SM:dress_fct}
\end{align}
Our choice is suggested by HTL results at high temperatures and small
coupling. In contrast to a non-interacting fermion, whose spectral
function consists of a single $\delta$-function, it is known from HTL
that an additional collective excitation  develops: the plasmino.
The fitting parameters $E_{1,2}$ denote the quasiparticle energies and
$Z_{1,2}$ the corresponding residues.
Certainly the full spectral function will be more complicated. The
working assumption here is, that if quasiparticle excitations with
small decay widths exist, then their peaks will be the dominant
contribution of the spectral function and therefore such a simple
ansatz may reveal characteristics of the quasiparticles and their dispersion relations.
However, we emphasize again that in HTL calculations the thermal masses
$E_{1/2}$ are of order $gT$, whereas the width of the corresponding
peaks in the spectral functions are of order $g^2T$. In a strong
coupling regime this becomes of the same order, if not larger.

Likewise as done in Ref.~\cite{Karsch:2009tp} we also investigated
other ans{\"a}tze for the spectral function. We considered a single
pole ansatz
\begin{align}
\rho^M_{\pm}(\omega)
&=
2\pi Z_1\delta(\omega\mp E_1)
\,,
\label{one_pole:delta} 
\end{align}
and ans{\"a}tze allowing for one- and two-particle excitations with
Gaussian widths
\begin{align}
\rho^M_\pm(\omega)&=
2\sqrt{\pi}\,\frac{Z_1}{\Gamma_1}\exp{\frac{-(\omega\mp
    E_1)^2}{\Gamma_1^2}}
\,,
\label{one_pole:width}
\\
\rho^M_\pm(\omega)
&=
2\sqrt{\pi}\,\bigg[
\frac{Z_1}{\Gamma_1}\exp{\frac{-(\omega\mp E_1)^2}{\Gamma_1^2}}
\nonumber\\ &
\phantom{2\sqrt{\pi}\,\bigg[+}
+
\frac{Z_2}{\Gamma_2}\exp{\frac{-(\omega\pm E_2)^2}{\Gamma_2^2}}
\bigg]
\,.
\label{two_pole:width}
\end{align}
Here $\Gamma_1$ and $\Gamma_2$ are additional fitting parameters.

In order to evaluate the quality of the fit we minimize
\begin{align}
\ell_\pm^2
&=
\sum_n^{N_\omega} \left\vert S^{{\rm DSE},\,M}_\pm(i\omega_n) -
  S^{M}_{\pm}(i\omega_n)  \right\vert^2
\,,
\label{eq:l2}
\end{align}
i.e. we implicitly assume uncorrelated data with equal total errors.
For comparing the different fit forms we evaluate $\ell^2_+$ with
$N_{\omega}=39$. In principle, one may also define 
$\ell_\pm^2/\rm{dof}$ where $\rm{dof}$ denotes $N_\omega$ minus the
number of fit parameters however this does not affect our results.

A remark on the numerical results seems appropriate here.
Within a given truncation scheme the error in Dyson-Schwinger
calculations is essentially determined by the error of the numerical integration. 
It can be estimated by varying the numerical parameters of the
integration and can be made significantly smaller than in lattice
calculations. 
At least for large temperatures where the Matsubara frequencies are
largely separated, the assumption of uncorrelated errors seems
reasonable. 
In principle, the self-energy is calculated in DSE calculations and the
total error appears for the inverse propagator, which grows linearly
in frequencies $\omega_n$. 
By using Eq.~(\ref{eq:l2}) we therefore enhance the importance of small
$\omega_n$. We expect the relevant information for spectral
functions to be encoded here as the propagator approaches the
perturbative result at large frequencies very fast.

For the data analyzed we find $\ell_+^2$ more than one order
of  magnitude smaller for fits based on Eq.~(\ref{two_pole:delta})
than for the single pole Ansatz Eq.~(\ref{one_pole:delta}).
This is in accordance with the findings in  Ref.~\cite{Karsch:2009tp}.
Comparing $\ell^2_+$ obtained from the single peak ansatz
Eq.~(\ref{one_pole:width}) with its value obtained for ansatz (\ref{two_pole:delta})
we find similar results for light quarks ($m/T\lesssim 0.2$)
but roughly one order of magnitude difference in favor
of (\ref{two_pole:delta}) for heavy quarks ($m/T\gtrsim 0.2$).

Examining ansatz (\ref{two_pole:width}) we find parameters with improved
$\ell_+^2$ compared to the two pole
ansatz. However depending on the initial parameter values different
local minima may be found.
Typical solutions correspond to a normal mode with a width which
tends to zero and a plasmino mode with broad width.
Furthermore we find that the possible excitations do not acquire a thermal
mass in the chiral limit.
In Ref.~\cite{Karsch:2009tp} it was found that ansatz (\ref{two_pole:width})
may improve $\chi^2/\rm{dof}$ for uncorrelated fits but reduces
to two delta functions for correlated fits.
By way of comparison and
taking into account the results of \cite{Karsch:2009tp} for correlated
data we focus the
discussion to the two-pole ansatz in the following.

\subsection{Results above and below the critical temperature}
Before analyzing the quark mass dependence of the fit parameters
we investigate the propagator below and above the critical
temperature. Here, $T_c$ is given by the deconfinement transition
as extracted from the lattice simulations of our input gluon
propagator and from the dressed Polyakov loop extracted from 
the quark propagator, see \cite{Fischer:2010fx} for details. The
chiral transition temperature coincides with $T_c$ within a few MeV,
depending slightly on the bare quark mass.

\begin{figure}
\includegraphics[width=.9 \columnwidth]{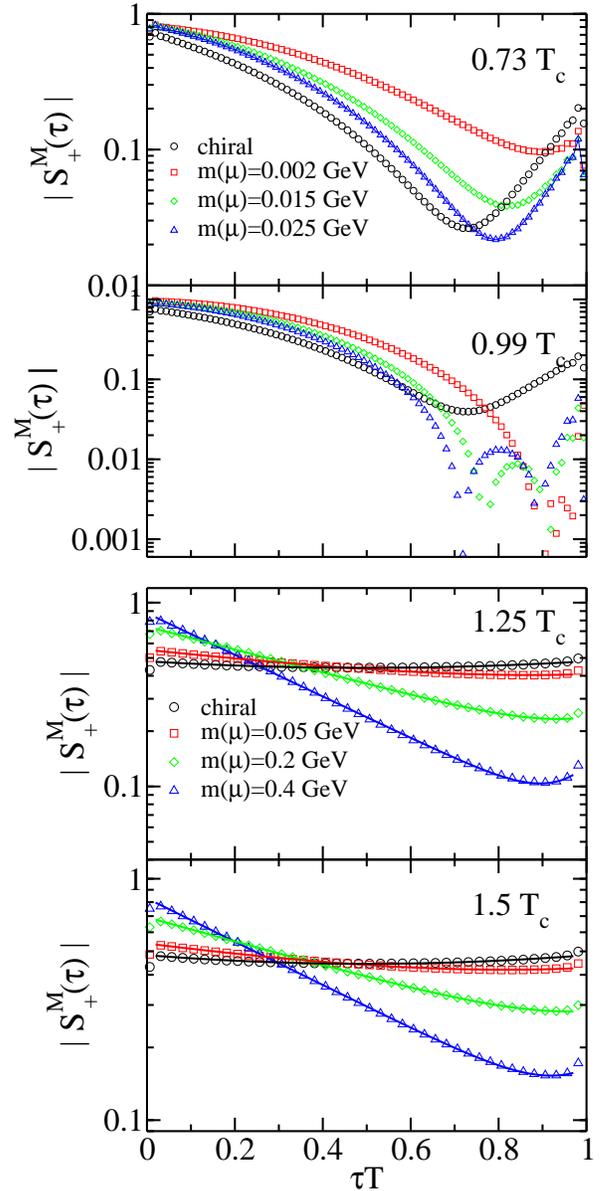}
\caption{Schwinger function $S^M_+(\tau)$ for different temperatures
  and current quark masses. For $T=0.99 T_c$ and
  $m(\mu)=0.2,\,0.4\,\rm{GeV}$ the Schwinger function changes sign in
  an interval around $\tau T\approx 0.8\rm$. 
 }
\label{fig:schwinger_fct}
\end{figure}

In Fig.~\ref{fig:schwinger_fct} we show the Schwinger function
defined by the Fourier transform of $S^M_+(i\omega_n)$:
\begin{align}
\label{eq:Stau}
S^M_+(\tau)
&=
-T\sum_n\,e^{-i\omega_n\tau}\,S^M_+(i\omega_n)
\,.
\end{align}
The numerical data are denoted by data points.
Above $T_c$ the solid lines are obtained by using the spectral
function obtained from fitting  with ansatz (\ref{two_pole:delta}).
Note that the points close to $\tau T=0$ and $\tau T=1$  are subject
to boundary effects due to the finite number of Matsubara modes used
when Fourier transforming.

At zero temperature the Schwinger function has been used to search for
positivity violations in the quark propagator. From the
Osterwalder-Schrader axioms of
Euclidean quantum field theory one knows that the Schwinger function
needs to be strictly positive for physical particles. Conversely, positivity
violations in the Schwinger function are a sufficient criterion for the 
absence of the corresponding particle from the physical spectrum of a theory.
Using the spectral representation (\ref{prop:spectral}) in Eq.~(\ref{eq:Stau}) 
it is also easy to see, that a positive spectral function always gives a 
positive Schwinger function. For the quark propagator at zero temperature 
these properties have been investigated in a number of works, 
see \cite{Alkofer:2003jj} and Refs. therein. At finite
temperature it has been suggested from calculations in a simple model, that 
positivity violations only occur below the critical temperature, whereas 
positivity is restored above $T_c$ \cite{Bender:1996bm}. Although this is not
exactly what we see in our more elaborate calculation, there are clear signals
for a qualitative change in the Schwinger function at $T_c$.
This is in line with quenched lattice QCD results that directly
determine $S^M_{+}(\tau)$~\cite{Karsch:2009tp}.

Above the critical temperature $S^M_+(\tau)$ is positive and found to
be convex on a log-scale.
Furthermore chiral symmetry restoration through $B(i\omega_n,0)=0$
translates into $S^M_+(i\omega_n)=-S^M_+(-i\omega_n)$ and
$S^M_+(\tau)=S^M_+(1/T-\tau)$.
We see this symmetry emerge when decreasing the current quark mass
$m(\mu)$.
Related to dynamical symmetry breaking this symmetry is absent below
$T_c$, even in the chiral limit.
More striking, however, is the concave shape around $\tau T \approx
0.5$ on a log-scale.
This cannot be reproduced by any ansatz including real poles only.
This finding again agrees with quenched lattice calculations in
Ref.~\cite{Karsch:2009tp}.

In addition we find positivity violations at $T=0.99 \, T_c$ for a
range in current quark masses, albeit this effect is not seen in the 
chiral limit and also for the smaller temperature. Nevertheless,
the concaveness of the Schwinger function below $T_c$ is generic in 
the sense that it is independent of variations in $d_1$. It can also 
not be attributed to the dynamical generation of quark masses: we find 
concave curvature below the deconfinement transition temperature $T_c$ 
also in the chiral limit and for values of $d_1$ that do not allow for 
spontaneous dynamical mass generation. It is therefore apparent, that
the curvature of the Schwinger function is related to quark 
confinement \cite{Bender:1996bm} and may serve as another source for
extracting the deconfinement transition temperature $T_c$ from the
quark propagator besides the dressed Polyakov loop 
\cite{Fischer:2009wc,Gattringer:2006ci,Braun:2009gm}.

\subsection{Quark mass dependence of fit parameters}

\begin{figure}
\includegraphics[width=.9 \columnwidth]{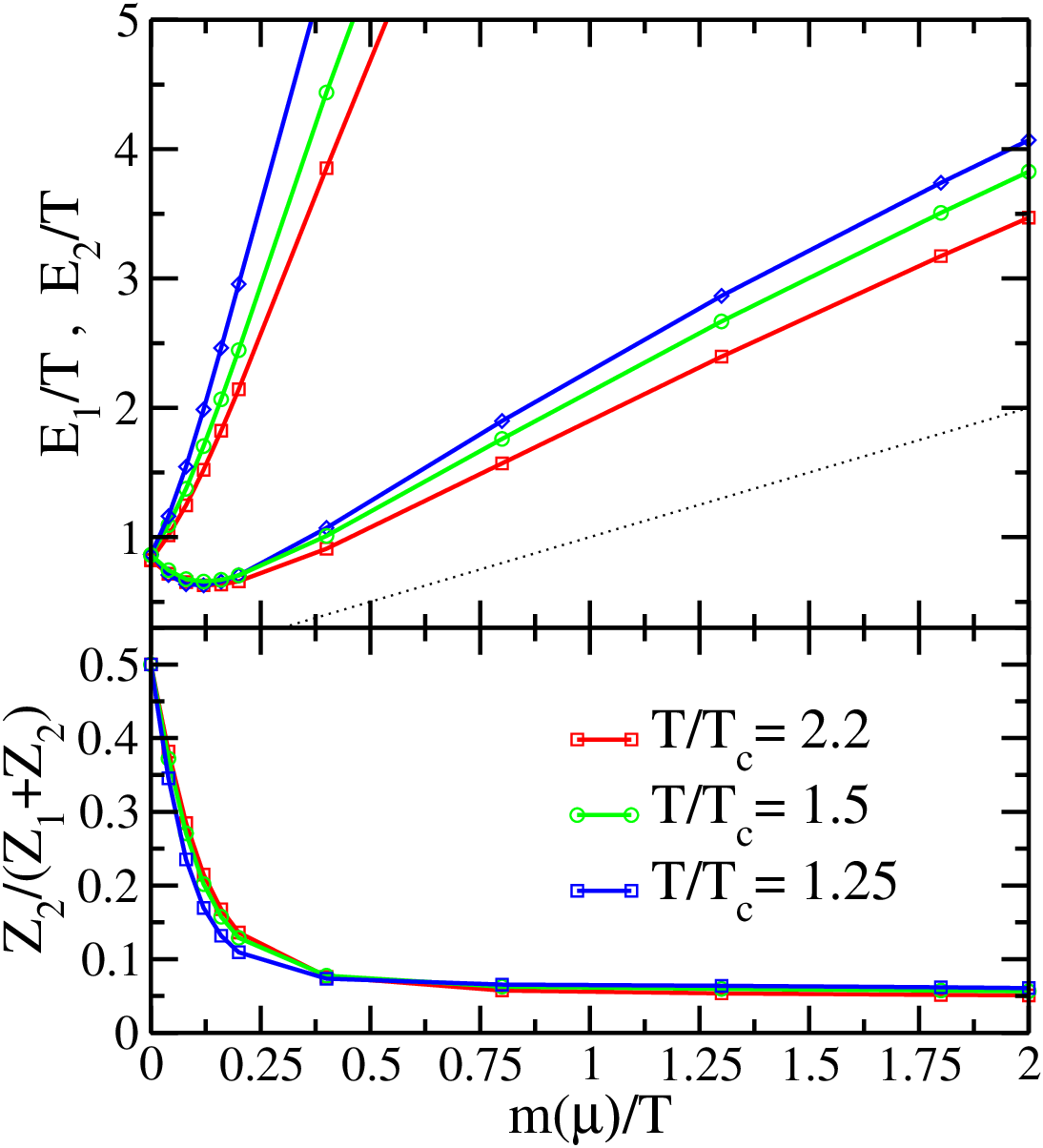}
\caption{Dependence of fit parameters in two-pole ansatz at vanishing
  momentum on current quark mass $m(\mu)$ for three temperatures. For
  the two branches in the upper plot we define $E_2\geq E_1$.
}
\label{fig:mass_dep}
\end{figure}

In the upper part of Fig.~\ref{fig:mass_dep} the dependence of the poles
$E_1$, $E_2$ on the current quark mass for three different temperatures
is shown. The lower part shows the relative strength of the 
plasmino pole, defined as the $E_2$ branch in the upper plot, which is
given by the ratio $Z_2/(Z_1+Z_2)$.
The disappearance of the scalar part of the propagator
in the chiral limit, i.e. $B(i\omega_n,0)=0$, manifests itself in the spectral function
being an even function,  $\rho(-\omega,\mathbf{0})=\rho(\omega,\mathbf{0})$, and therefore $E_1=E_2$, $Z_1=Z_2$.
We use this as the definition of the thermal mass $m_T\equiv E_1$ and,
as outlined in the context of Tab.~\ref{tab1}, use this value to
determine the vertex parameter $d_1$.
It is worth noting that the qualitative behavior in
Fig.~\ref{fig:mass_dep} does not change when increasing $d_1$ by one
order of magnitude.

From the mass dependence of $Z_2/(Z_1+Z_2)$ in Fig.~\ref{fig:mass_dep}
it is discernible that the plasmino contribution decreases with
increasing quark mass. This is in accordance with the
expectation that for heavy quarks the spectral function reduces
to one of free quarks. 
Only small deviations from sum rule (\ref{eq:sum_rule1}) are found
with a mismatch of less than $4\%$.
Furthermore we find the minimum of $E_1$ at  
non-vanishing quark mass whereas the plasmino pole increases
monotonically with quark mass. The slope of $E_1$ and $E_2$ for fixed
$m(\mu)/T$ is temperature dependent and decreases with increasing
temperature.
Compared to the lattice data in Ref.~\cite{Karsch:2009tp} we find
qualitative and quantitative similar results. The minima in $E_1$
approximately coincide with the lattice results whereas the slopes 
in $Z_2/(Z_1+Z_2)$ and $E_{1,2}$ are slightly different. However
this may be traced back to different definitions of the bare quark mass.

\section{Momentum dependence of the spectral function in the chiral limit}
\label{sec:mom_dep}

In this section we investigate the quark propagator in the chiral
limit for varying momenta.
For a chirally symmetric quark propagator it is convenient to use the
decomposition on energy projectors as shown in Eqs.~(\ref{eq:rhoP})
and (\ref{eq:SP}). In terms of the quark dressing functions we then
have
\begin{align}
S^P_{\pm}(i\omega_n,\vert\pp\vert)
&=
-
\frac{
i\omega_n C(i\omega_n,\vert\pp\vert)\pm \vert\pp\vert A(i\omega_n,\vert\pp\vert)
}{
\omega_n^2 C^2(i\omega_n,\vert\pp\vert)+\vert\pp\vert^2 A^2(i\omega_n ,\vert\pp\vert)}
\,.
\end{align}
Along the same line as in the previous section we use the two-pole ansatz
\begin{align}
\rho^P_\pm(\omega,\vert\pp\vert)
&=
2\pi\big[Z_1\delta(\omega\mp E_1)+Z_2\delta(\omega\pm E_2)\big]
\label{two_pole2:delta}
\end{align}
and fit the parameters $Z_{1}$, $Z_2$ and $E_{1}$, $E_2$ to the data.
These are then momentum dependent and we obtain the dispersion
relation of quasiparticle and plasmino together with there spectral
strength.
For all momenta shown here $\ell_+^2$ is of the same order of
magnitude than the results for zero momentum.

\begin{figure}
\includegraphics[width=.8 \columnwidth]{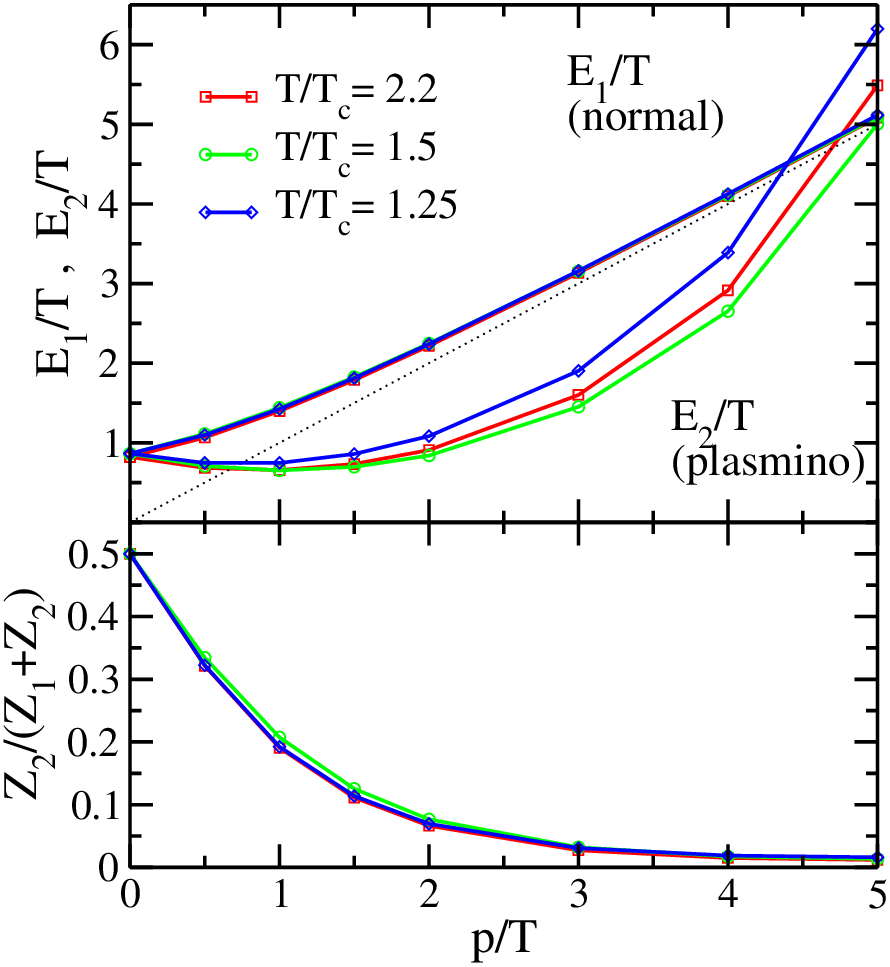}
\caption{Momentum dependence of the fit parameters for the two pole
  ansatz (\ref{two_pole2:delta}) at three different temperatures.}
\label{fig:disp_rel}
\end{figure}

In Fig.~\ref{fig:disp_rel} we show the results for the momentum
dependence of the parameters $E_1$, $E_2$ and $Z_2/(Z_1+Z_2)$
for temperatures $T/T_c=1.25$,
$1.5$ and $2.2$. For increasing momenta the ratio $Z_2/(Z_1+Z_2)$
decreases and tends to go to zero for high momenta.
Likewise $E_1$ is approaching the light cone. Therefore the spectral
function at high momenta reduces to the one of a single free particle.
In the plasmino branch a minimum at non-vanishing momentum is clearly 
recognizable.
This feature is known
from perturbation theory and may be anticipated from the results in
Ref.~\cite{Karsch:2009tp}. It is conspicuous and not obtained in one loop
perturbation theory that the plasmino branch goes over into the 
space-like region. However it is known that due to Landau damping
the spectral function at high temperatures contains a contribution in
the space-like region which is not included in our fit ansatz 
Eq.~(\ref{two_pole2:delta}). It might be that the plasmino pole by
shifting into the space-like region mimics this missing contribution.
This possibility is also considered in the following.
An analytic continuation of the DSE to include smaller Euclidean
energies might help to distinguish these different scenarios.
Also note that with increasing momenta, the value of $E_2$ becomes
less constrained by the fitting procedure, since the spectral strength
$Z_2$ and therefore the contribution of the plasmino branch to the data decreases.

\begin{figure}
\includegraphics[width=.8 \columnwidth]{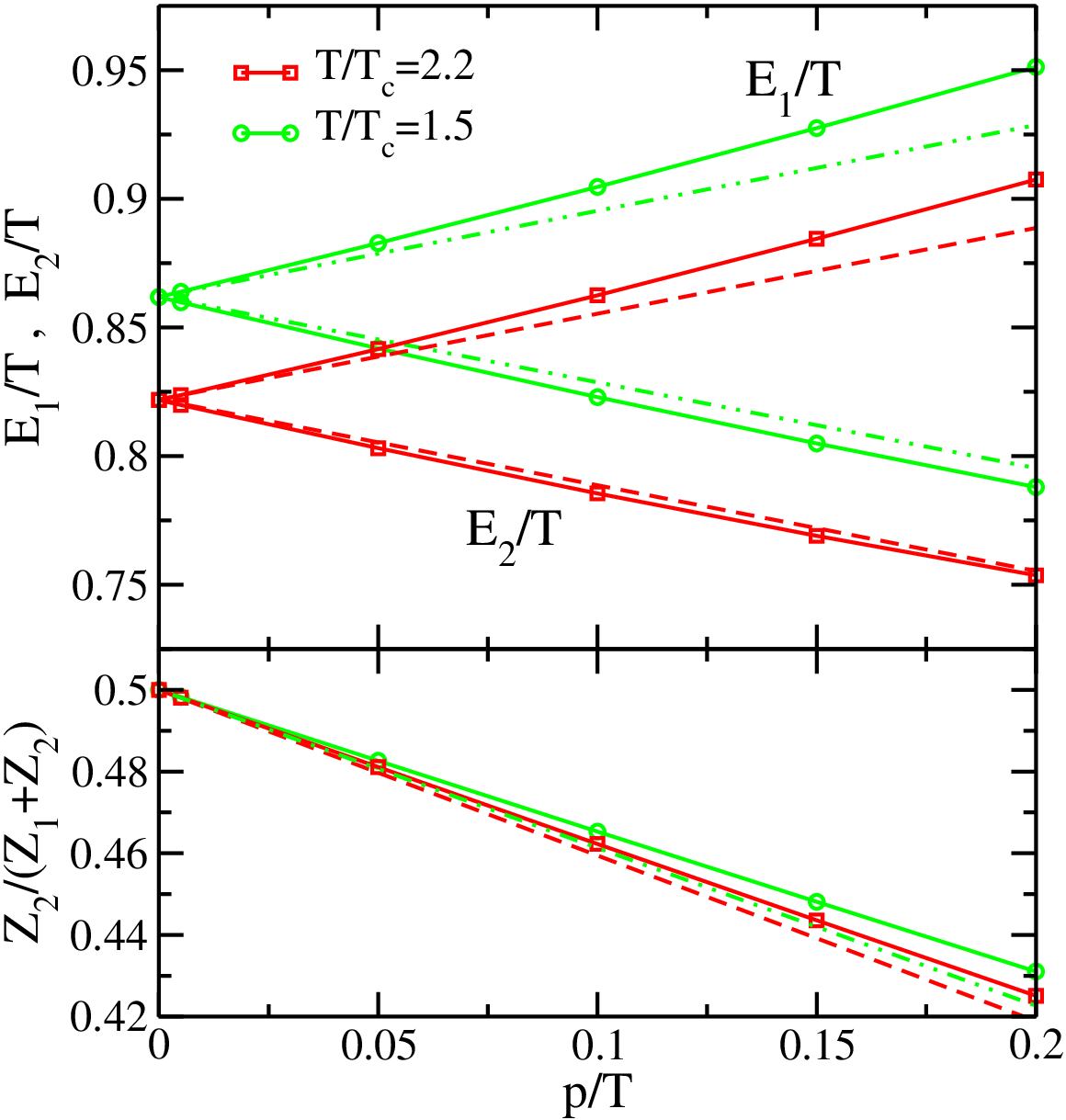}
\caption{Close up of Fig.~\ref{fig:disp_rel} in momentum range $p/T\in [0, 0.2]$.
For comparison we give the small momentum behavior of the HTL
approximation in Eqs.(\ref{eq:HTL_E}) and (\ref{eq:HTL_Z}) as dashed lines.}
\label{fig:disp_rel_close_up}
\end{figure}

In Fig.~\ref{fig:disp_rel_close_up} we focus on the region near
zero momentum. In addition to the fitted values shown as straight lines,
we also present the behavior for the HTL result~\cite{Bellac:1996} given by
\begin{align}
E_1\simeq m_T+\frac{\vert\pp\vert}{3}
\,,
&\quad
E_2\simeq m_T-\frac{\vert\pp\vert}{3}
\,,
\label{eq:HTL_E}\\
\frac{Z_2}{Z_1+Z_2}
&\simeq
\frac{1}{2}-\frac{\vert\pp\vert}{3 m_T}
\label{eq:HTL_Z}
\end{align}
for $\vert\pp\vert\ll m_T$.
We observe that our results agree well with the HTL result for sufficiently
small momentum, with the better agreement for the plasmino branch $E_2$, and then depart from it.
Since the expansion is supposed to apply for small $\vert\pp\vert/T$,
this is expected.
Comparing the two temperatures it is recognizable that the results at
higher temperature agree better with the HTL result, although we are
certainly not in a regime where we expect the HTL approximation to work.

\begin{figure}
\includegraphics[width=.8 \columnwidth]{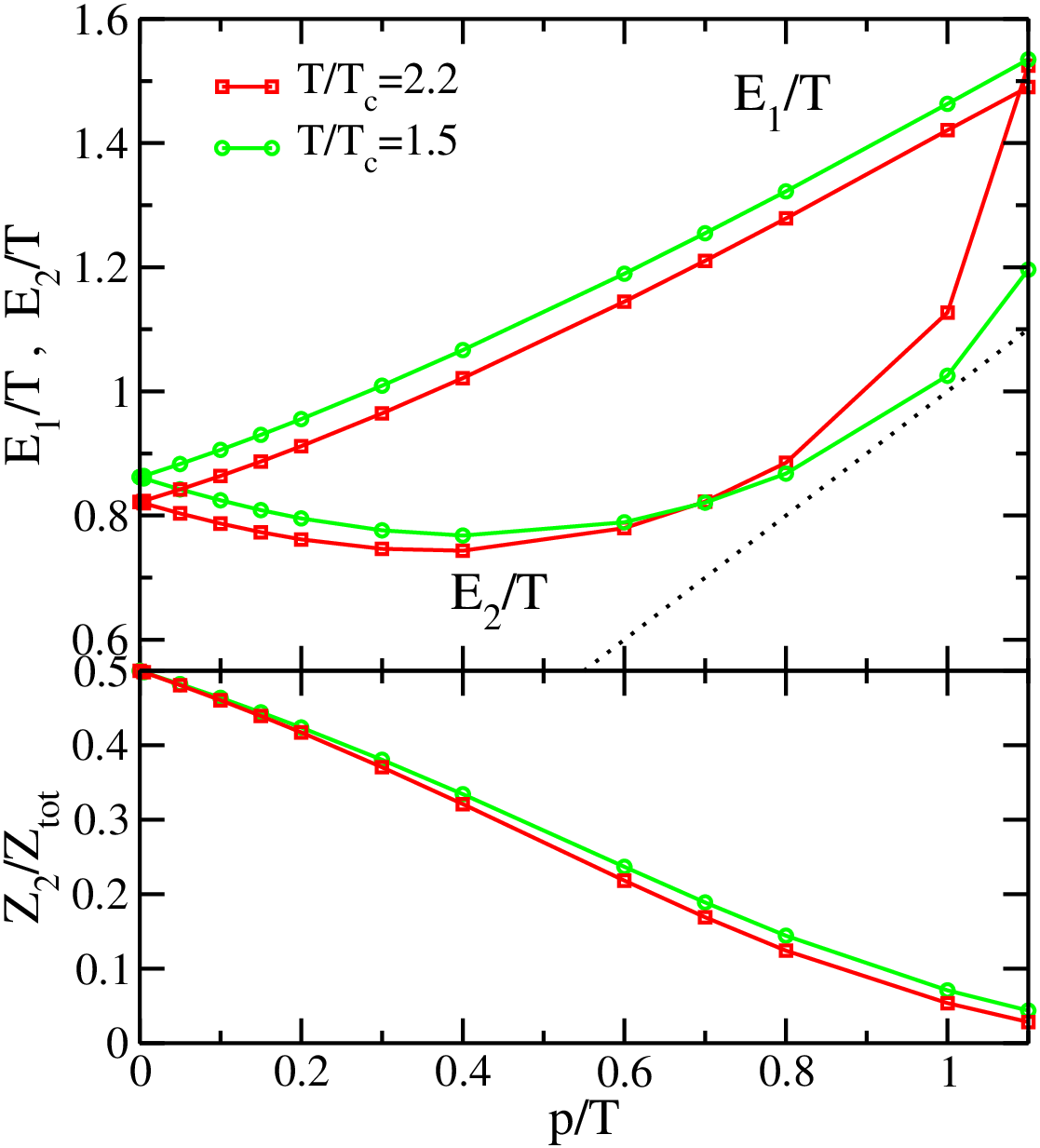}
\caption{Momentum dependence of the fit parameters for the extended
  ansatz (\ref{two_pole:delta_spacelike}) for two different
  temperatures.
}\label{fig:disp_rel_spacelike}
\end{figure}

Since the result for the plasmino branch of the two-pole ansatz
indicate a possible space-like continuum contribution we extend
the two-pole ansatz (\ref{two_pole2:delta}) by the HTL
continuum contribution.
The fitting function then matches exactly the spectral function in HTL approximation
\begin{align}\nonumber
\rho^P_\pm(\omega,\vpp&)
=
2\pi\big[
Z_1\delta(\omega\mp E_1)
+
Z_2\delta(\omega\pm E_2)\big]
\\ \nonumber
&+\;\frac{\pi}{\vpp}\,m_T^2\,(1\mp x)\Theta(1-x^2)\\ \nonumber
&\times\bigg[\Big(\vpp\,(1\mp x)\pm\frac{m_T^2}{2\,\vpp}\Big[(1\mp x)\\
&\times\ln\left\vert\frac{ x+1}{ x-1}\right\vert\pm 2\Big]\Big)^{2}
+\frac{\pi^2 m_T^4}{4\vpp^2}(1\mp x)^2\bigg]^{-1}
\label{two_pole:delta_spacelike}
\end{align}
where $x=\omega/\vpp$. The thermal masses are taken from Tab.~\ref{tab1}
and $E_{1,2}$ and $Z_{1,\,2}$ are the fit parameters. 
The results are shown in Fig.~\ref{fig:disp_rel_spacelike}.
We find $\ell_+^2$ to be marginally smaller compared to the
two-pole ansatz and we conclude from the results presented in
Fig.~\ref{fig:disp_rel}, that the plasmino branch going over into the
space-like region might well be an artifact of a missing space-like
spectral weight in ansatz (\ref{two_pole:delta}).
The upper plot of Fig.~\ref{fig:disp_rel_spacelike} shows that the
plasmino tends to the light cone for momenta $p/T\simeq 0.8$ instead
of crossing it.
Also the minimum of the plasmino branch has shifted to lower
momenta.
At high momenta the plasmino seems to deviate from the light cone again.
However we note again that due to the small spectral strength $Z_2$ at these momenta
the value of $E_2$ is only poorly constrained.
 On the other hand the momentum dependence of $E_1$ is
practically unaltered.

The lower plot of Fig.~\ref{fig:disp_rel_spacelike} shows the relative strength
of the plasmino $Z_2/Z_{\rm{tot}}$ where
\be
Z_{\rm{tot}}=Z_1+Z_2+\int^{+\vpp}_{-\vpp}\!d\omega\,\rho^P_{\pm}(\omega,\vpp)\,.
\ee
Compared to the two-pole ansatz the plasmino spectral weight decreases considerably
more rapid with increasing momentum.
We also find that $Z_{\rm{tot}}$ agrees within $3\%$ with the
expected value obtained from sum rule (\ref{eq:sum_rule1}).
Since the spacelike part is subleading for small momenta the small
momentum behavior of $E_{1,2}$ and $Z_2/(Z_1+Z_2)$ is practically unaltered
by the continuum contribution.

\section{Quark spectral functions via MEM \label{sec:MEM}}

In order to shed more light on the possible shape of the spectral
functions and also on the truncation for gluon propagator and
quark-gluon vertex\footnote{Note that their parameterization used here
  is only constrained in Euclidean space, but their analytic
  continuation is in principle relevant for the shape of the quark
  spectral function.}, we employ MEM.
For simplicity our discussion will be based on historical MEM (see
e.g. Refs.~\cite{Jarrell:1996,nr:1997}), where we maximize the
functional
\begin{align}
Q[\rho]&=\alpha S[\rho] - \frac{1}{2}\chi^2[\rho]
\end{align}
in the spectral function $\rho$.
Its constituents are the entropy $S[\rho]$ and the
$\chi^2[\rho]$-distribution.
In the case of a scalar channel as in Eq.~(\ref{prop:spectral}) and
Eq.~(\ref{prop:spectralP}) it is common to use the Shannon-Jaynes
entropy
\begin{align}
S[\rho_\pm]
&=
\int d\omega \left(
\rho_\pm(\omega)-m(\omega)-\rho_\pm(\omega)\log\left(\frac{\rho_\pm(\omega)}{m(\omega)}\right)
\right)
\end{align}
for the entropy functional with prior estimate $m(\omega)$.
Features of this choice are that there is only one minima in $Q[\rho]$
for each choice of $m(\omega)$ and that $\rho(\omega)$ has the same
sign as $m(\omega)$.
For the $\chi^2[\rho]$-distribution between the propagator given
through $\rho(\omega)$ and the data from DSE calculations we will
assume
\begin{align}
\chi^2[\rho_\pm]
&=
\frac{1}{\sigma^2}\ell_\pm^2
\,.
\end{align}
As discussed before this corresponds to assuming uncorrelated data
with equal errors and we will need to specify the variance
$\sigma^2$.
For the prior estimate, which essentially determines the support of
the spectral function and its semi-positivity, we assume
\begin{align}
m(\omega)
&=
m_0\,\theta(\Lambda^2-\omega^2)
\,.
\end{align}
The magnitude $m_0$ and the cutoff $\Lambda$ will be varied in the
following to analyze their influence on the shape of the extracted
spectral function.
The method as presented here follows the philosophy of
regularization~\cite{nr:1997}: Since the inversion of the spectral
representation is an ill-posed problem, we want to impose the
positivity condition on the spectral function in a way that leads to a
well-defined problem.

In historical MEM the Lagrange multiplier $\alpha$ is then chosen such
that $\chi^2[\rho_\pm]=N_\omega$.
Within our setup we  observe, however,  that the extracted spectral
functions $\rho_\pm(\omega)$ actually only depends on the combination
$\bar{\alpha}=\alpha \sigma^2$ and that in this case
$\sigma^2=\ell_\pm^2/N_\omega$.
We will exploit this feature by varying $\bar{\alpha}$ in the
following, in particular because we do not want to give an estimate
for $\sigma^2$ at this point.

\begin{figure}
\begin{center}
\includegraphics[width=.44\textwidth]{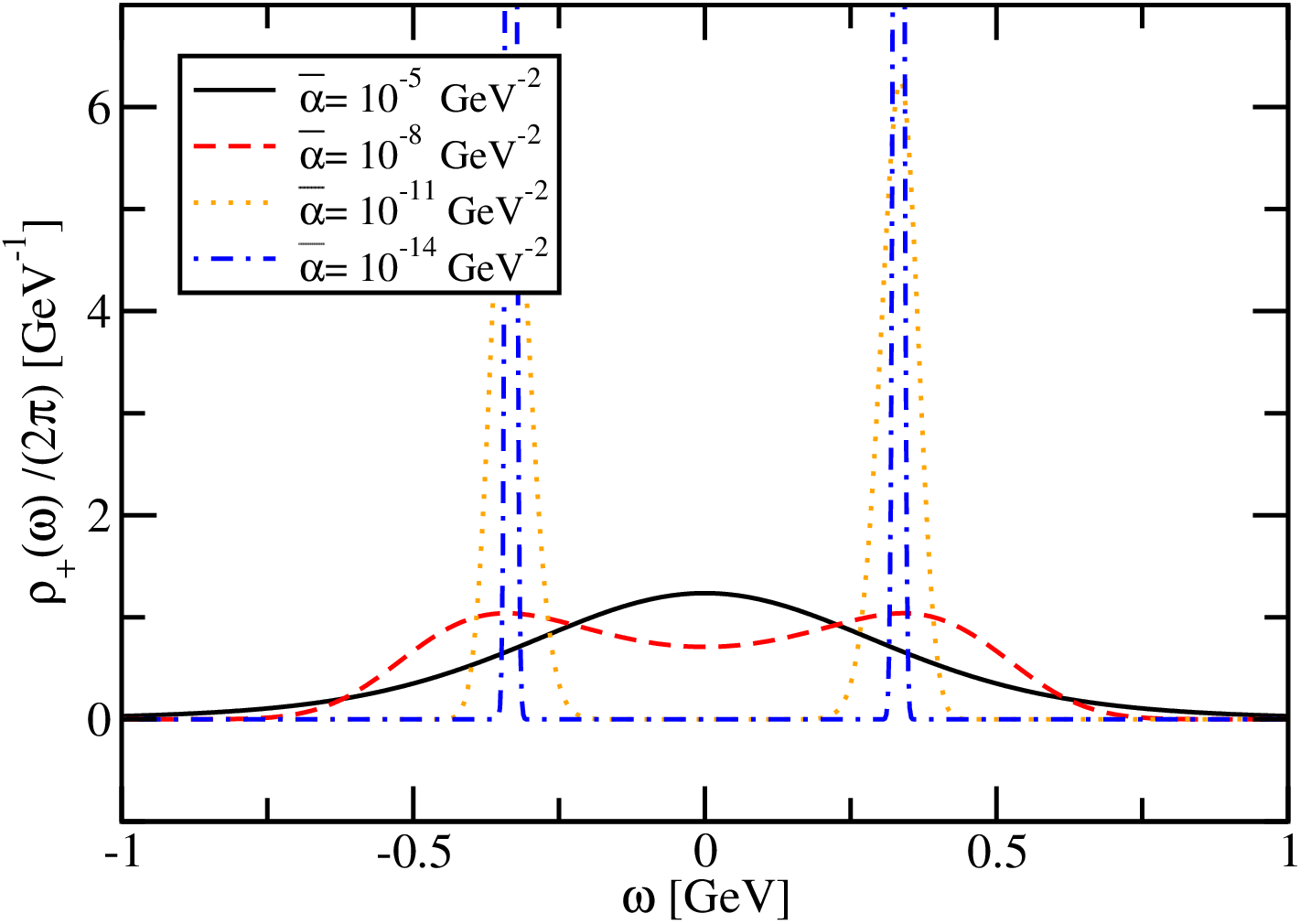}
\includegraphics[width=.48\textwidth]{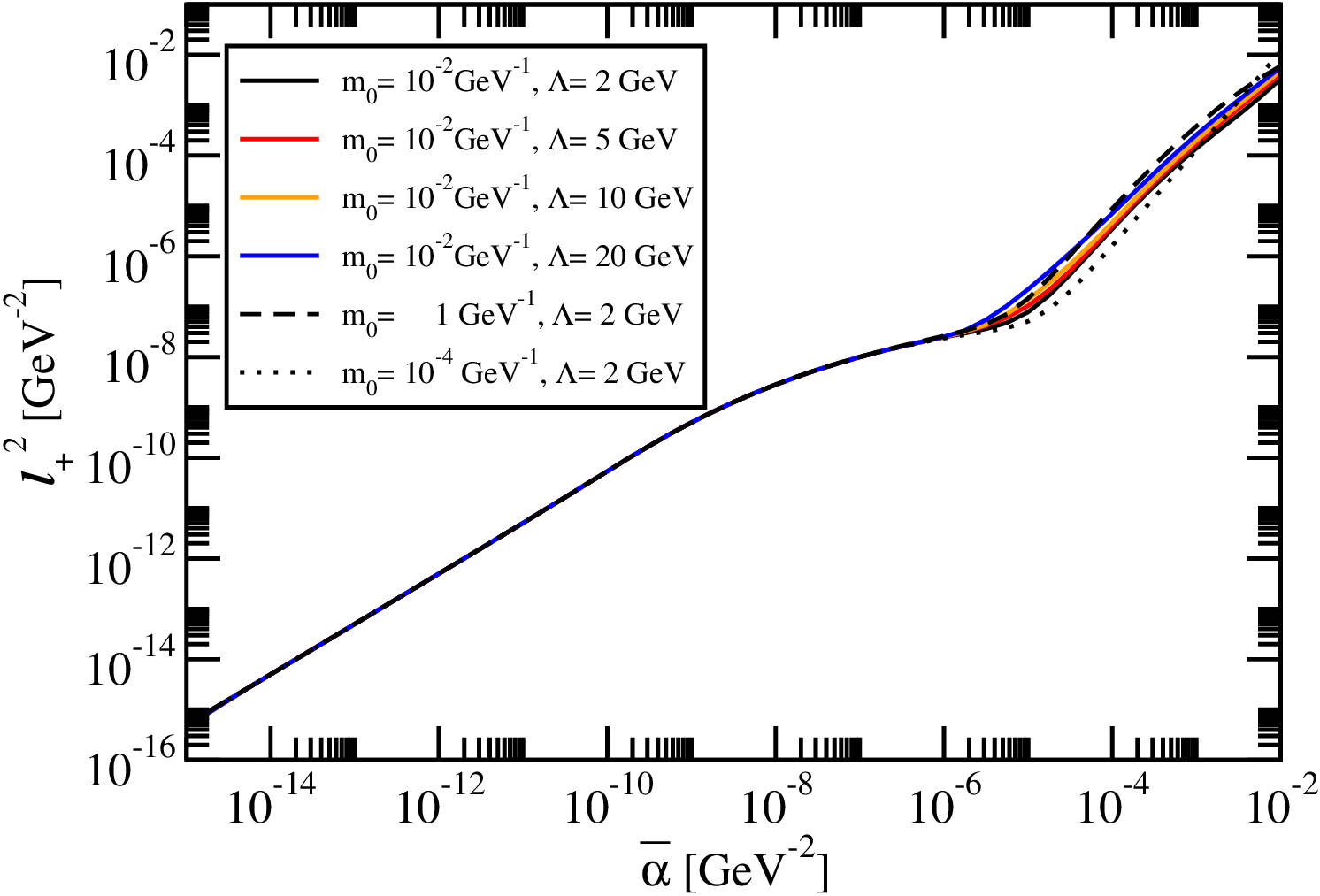}
\caption{
Upper panel: Spectrals function extracted from 'mock' data at various
values of $\bar{\alpha}$ for $m_0=10^{-2} {\rm GeV}^{-1}$ and
$\Lambda=2{\rm GeV}$.
Lower panel: $\ell_+^2$ as function of $\bar{\alpha}$ for various $m_0$
and $\Lambda$.
For both plots we have $T=1.5T_c$, $M=0.8\times1.5 T_c$ and
$N_\omega=20$.
}
\label{fig:MEM_HTL}
\end{center}
\end{figure}

In order to get an intuition for the way MEM works, we consider
$\pp=0$ and produce 'mock data' by determining $S_+(\omega_n)$ for
the spectral function
\begin{align}
\rho_+(\omega)
&=
2\pi\left(\frac{1}{2}\delta(M-\omega)+\frac{1}{2}\delta(M+\omega)\right)
\end{align}
analytically.
This corresponds to the two-pole ansatz~(\ref{two_pole:delta}) and
also with the HTL result at $\pp=0$.

Our results are summarized in Fig.~\ref{fig:MEM_HTL}. By the example
of $m_0= 10^{-2} {\rm GeV}^{-1}$ and $\Lambda=2\,{\rm GeV}$, we observe
that the correct spectral function is reproduced when making
$\bar{\alpha}$ sufficiently small. This is very generic, in particular
we checked that the extracted spectral function for a given
$\bar{\alpha}\in [10^{-15}{\rm GeV}^{-2}, 10^{-5}{\rm GeV}^{-2}]$ is
indistinguishable on the plot when varying $m_0\in[1\,{\rm
  GeV}^{-1},10^{-4}{\rm GeV}^{-1}]$ and $\Lambda\in[2\,{\rm GeV}, 20\,
{\rm GeV}]$.
For the corresponding $\ell_+^2$, which should be equal to $N_\omega \sigma^2$ by choosing an appropriate $\alpha$,
 we find on the one hand side, that the
dependence on the prior estimate is most extreme at larger
$\bar{\alpha}$, albeit in general the sensitivity can be considered as
mild.
More importantly, we find that $\ell_+^2$ for different prior estimates
converges quickly upon decreasing $\bar{\alpha}$ and that it converges
towards zero.
This can be qualitatively understood from the fact that we know that
there is a positive definite spectral function which reproduces the
data and that there is a competition between $\alpha S[\rho_+]$ and
$\frac{1}{2}\chi^2[\rho_+]$ when maximizing the functional $Q[\rho_+]$.
This leads to $\alpha S[\rho_+]\sim\frac{1}{2}\chi^2[\rho_+]$ and
$S[\rho_+]\to{\rm const}>0$.
 Consequently, $\ell_+^2$ should go to zero when taking
 $\bar{\alpha}\to0$.

\begin{figure}
\begin{center}
\includegraphics[width=.48\textwidth]{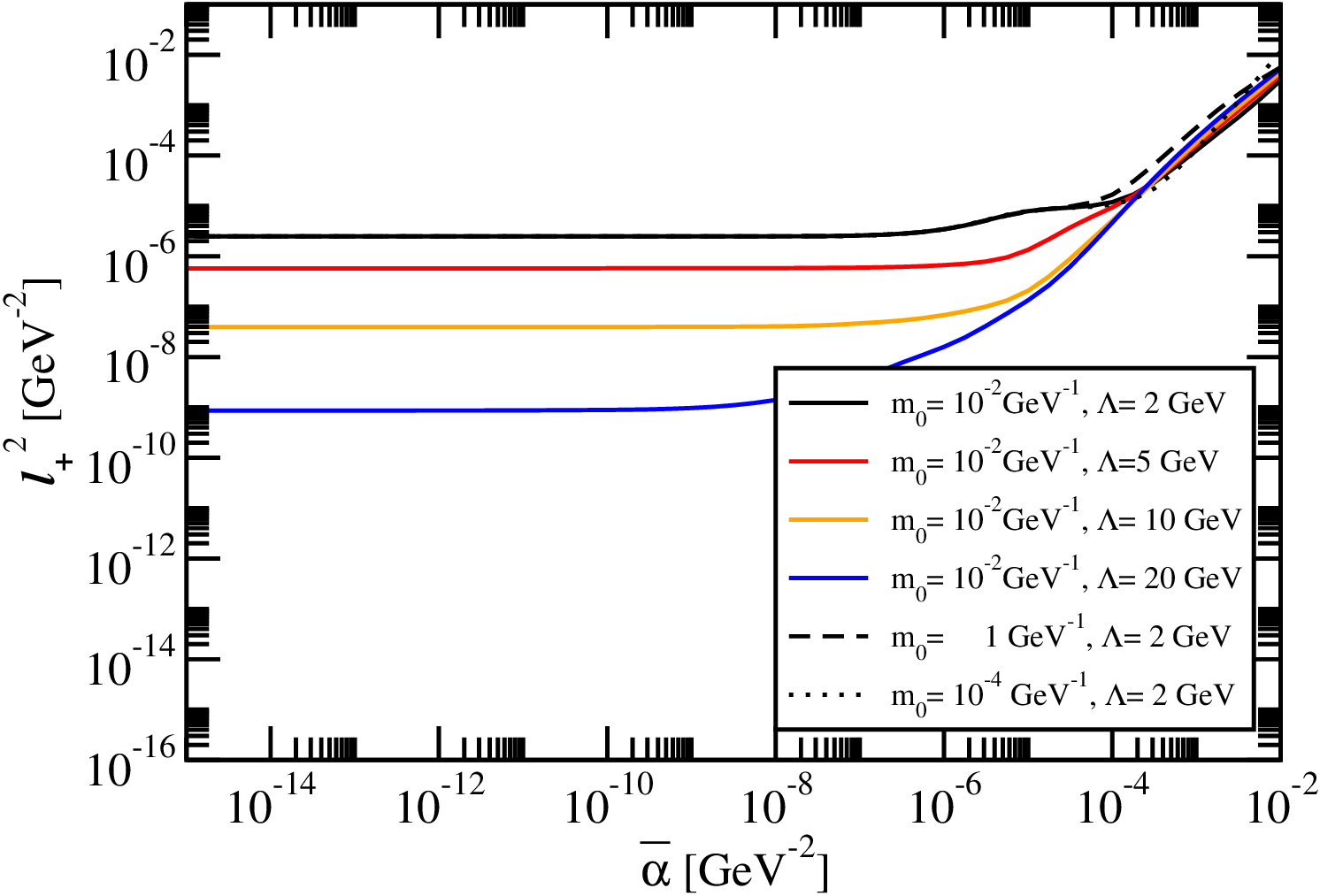}
\caption{
$\ell_+^2$ between numerical DSE results and $S_+(\omega_n)$ for MEM
extracted spectral function as a function of $\bar{\alpha}$ for
various $m_0$ and $\Lambda$ as well as $T=1.5\,T_c$  and $N_\omega=20$.
}
\label{fig:MEM_DSE1}
\end{center}
\end{figure}

\begin{figure}
\begin{center}
\includegraphics[width=.44\textwidth]{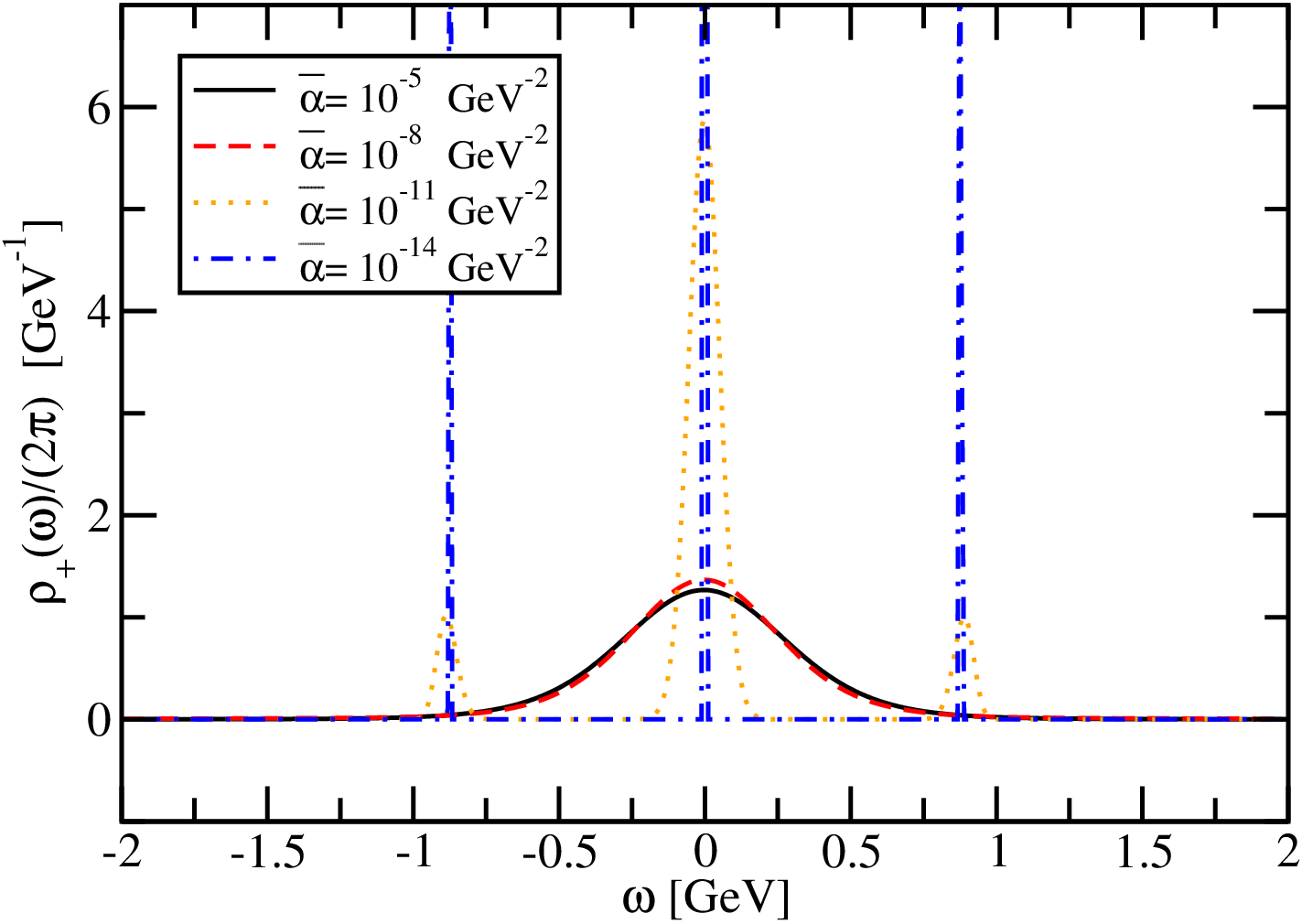}
\includegraphics[width=.44\textwidth]{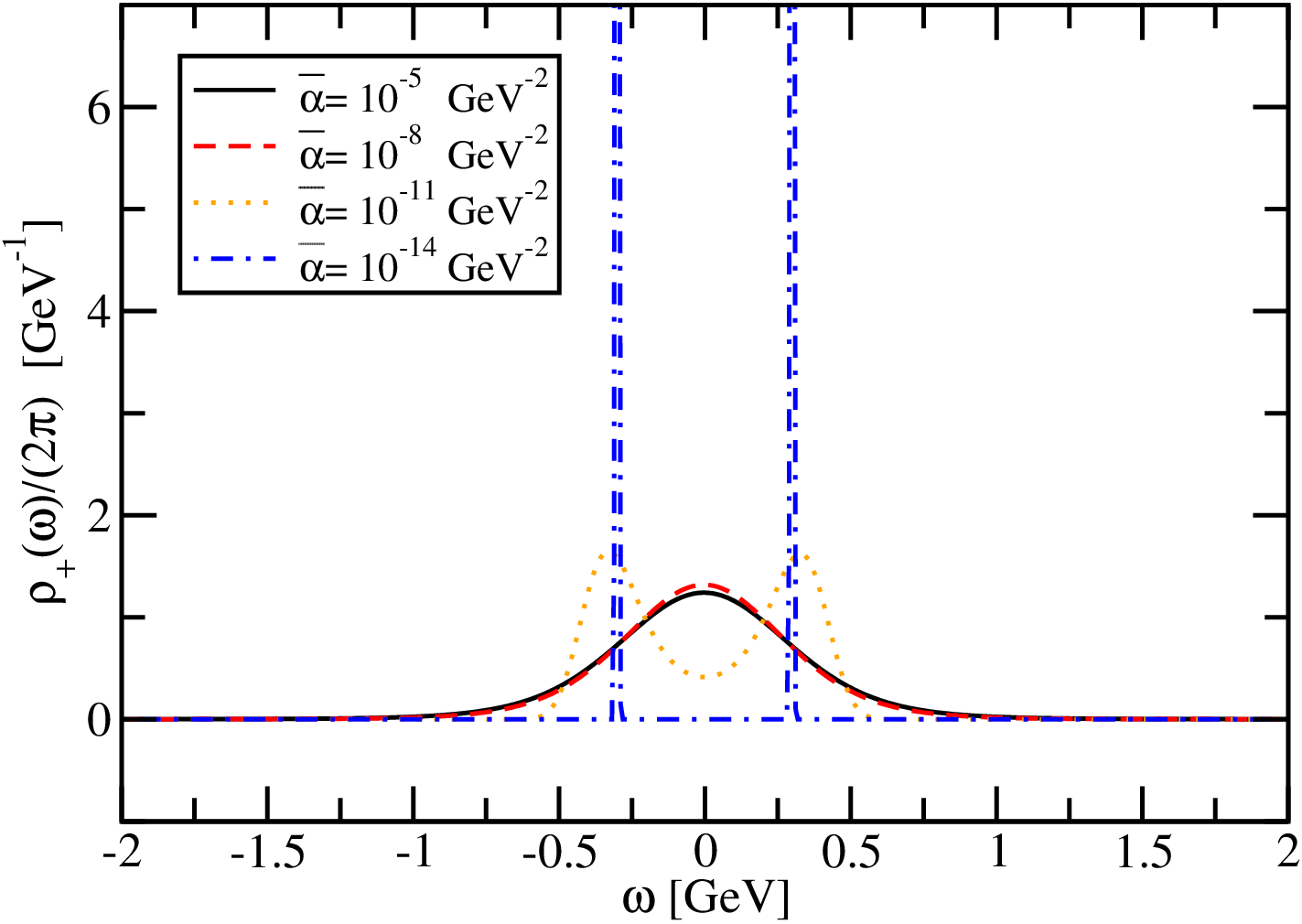}
\caption{
Upper panel: Spectral functions extracted from numerical DSE results
at various values of $\bar{\alpha}$ for $m_0=10^{-2} {\rm GeV}^{-1}$,
$\Lambda=10\,{\rm GeV}$ and  $N_\omega=20$.
Lower panel: Same us upper plot but for  $\Lambda=20\,{\rm GeV}$.
}
\label{fig:MEM_DSE2}
\end{center}
\end{figure}

Before discussing errors on our computed DSE 'data', we want to
present the results when applying MEM to our results in a similar
fashion as discussed above.
For this purpose we present $\ell_+^2$ as a function of $\bar{\alpha}$
at $\pp=0$ for various prior estimates in Fig.~\ref{fig:MEM_DSE1} and
illustrate some extracted spectral functions for $m_0=10^{-2} {\rm
  GeV}^{-1}$ as well as $\Lambda=10\,{\rm GeV}$ and $\Lambda=20\,{\rm
  GeV}$ in Fig.~\ref{fig:MEM_DSE2}.

The two-pole fit ansatz (\ref{two_pole:delta}) for this specific case
yields $\ell_+^2\approx10^{-5}{\rm GeV}^{-2}$. As anticipated we obtain
better fitting spectral functions for small enough $\bar{\alpha}$,
here $\bar{\alpha}\lesssim10^{-5}{\rm GeV}^{-2}$ for the considered
prior estimates $m_0\in[1{\rm GeV}^{-1},10^{-4}{\rm GeV}^{-1}]$ and
$\Lambda\in[2\,{\rm GeV}, 20\, {\rm GeV}]$. As before we also find that
the dependence on $m_0$ is very weak and we can therefore focus on the
spectral function's support given through $\Lambda$ only.
It is, however, striking that the $\ell_+^2$ has a strong dependence on
$\Lambda$. Of course we should be able to obtain better fitting
spectral function when increasing the support, the results suggest
however that a few percent of the spectral strength come from energies $\vert\omega\vert>2{\rm GeV}$
near $\Lambda$. This is unexpected.
As shown in Fig.~\ref{fig:MEM_DSE2} this does not affect the shape
of the extracted spectral function in the domain $\omega\in[-2\,{\rm
  GeV},2\,{\rm GeV}]$ for $\bar{\alpha}\approx 10^{-8}{\rm GeV}^{-2}$
noticeably. In case of the mock data we could already observe a
two-peak structure at this point.

When decreasing $\bar{\alpha}$ beyond $10^{-8}{\rm GeV}^{-2}$, we find
a strong dependence on $\Lambda$ and even more puzzling a fast
saturation of $\ell_+^2$.
The fact that the spectral functions tend towards a sum of peaked
functions should be considered a generic feature of method and not be
over-interpreted. The agreement at $\Lambda=20{\rm GeV}$ with the
two-pole ansatz is probably a coincidence, in particular since the
true spectral function is expected to have a finite width.

The reason for our conservatism is the value of $\ell_+^2$ that can be
obtained within MEM. Even for an allowed support of $[-20{\rm
  GeV},20{\rm GeV}]$, we cannot obtain $\ell_+^2\leq 8\,\, 10^{-10}{\rm
  GeV}^{-2}$.
 On the other hand a variation of numerical parameters
suggests that the numerical error on our 'data' is in fact below this
value. Therefore, strictly speaking, historical MEM requiring
$\chi^2=N_\omega$ cannot be applied.

A plausible explanation for this finding would be a spectral function
that, at least for the applied truncation scheme, is not positive
semi-definite.
To elaborate this a bit further, it is known e.g. from the calculation
in Ref.~\cite{Harada:2007gg}, that the determination of the spectral
functions via the quark's DSE requires knowledge of gluon-propagator
and quark-gluon vertex in the time-like domain and in particular of
the gluon's spectral function.
For the parameterization of the gluon used in this work, it can be
checked that its spectral function is not positive semi-definite.

We therefore consider the analysis via MEM on the one hand side as an
advise for caution when looking at the fits discussed in the earlier
sections. On the other hand it motivates future work for the direct
determination of the spectral function and a better analysis of the
truncation scheme.

\section{Conclusions and outlook \label{sec:sum}}

In the present work we have analyzed the quark propagator at finite
temperatures obtained within a recently employed truncation of the
quark's Dyson-Schwinger equation.
The investigation was in part motivated by lattice QCD calculations of
the quark propagator in Landau gauge using the quenched approximation,
which - in addition - also attempted to extract the spectral function
of the propagator from the obtained data.

For the Matsubara propagator, i.e. in Euclidean space,  we found
qualitative agreement with the lattice QCD results below and above the deconfinement
phase transition.
General analytic properties of the quark propagator are different
below and above the deconfinement phase transition temperature
and as our results indicate, determined by the non-perturbative behavior of the gluon propagator.
Consequently, we also conclude that our numerical data below the
deconfinement phase transition do not allow for an interpretation in terms of
a quasi-particle picture.
Above the deconfinement phase transition, on the other hand, we could fit the
data by a physically motivated spectral function consisting of a
quasiparticle branch and a plasmino branch with vanishing width,
respectively.
The obtained results for the current quark mass dependence of the
thermal masses and for the dispersion relations of quasiparticle and
plasmino are then consistent with those from quenched lattice QCD.
Furthermore we included a possible spacelike continuum contribution
in the model spectral function which was not considered in the lattice calculations.
We find this contribution to shift the dispersion relation of the plasmino
into the timelike region. This indicates the importance
of additional continuum contributions in the spectral function.

In addition we also analyzed our results by means of the so-called
historical maximum entropy method. Although the method was shown to
work for 'mock' data once the input was assumed to be accurate enough,
the application to the numerical DSE results turned out to be more
subtle.
We interpreted the results as a hint that the spectral function, at
least in the employed truncation, might not be positive
semi-definite.

In summary, we conclude from this that we can achieve nice agreement
with lattice QCD data in Euclidean space, but extracted spectral
functions using fit forms should be taken with a grain of salt.
Within the framework of Dyson-Schwinger equations this motivates
future work of directly evaluating the spectral function. Although
technically more evolved, calculations within models or using simpler
truncations indicate that this is doable.
Also the sensitivity on the gluon propagator and the quark-gluon
vertex should be investigated, in particular since their analytic
continuation to Minkowski space and therefore their spectral function
is relevant for the outlined calculation.
Furthermore it might be illustrative to analyze the lattice results of
the gluon propagator along similar lines as has been done for the results of the quark
propagator in quenched approximation.

\section*{Acknowledgement}
C. F. and J. M. were supported by the Helmholtz Young Investigator
Grant VH-NG-332, by the Helmholtz Alliance HA216-TUD/EMMI and the
Helmholtz International Center for FAIR within the LOEWE program of
the State of Hesse.
D.N. was supported by the German Research Foundation (DFG) under grant
number Ni 1191/1-1 and by the DOE Office of Nuclear Physics under
grant DE-FG02-00ER41132.


\end{document}